\RequirePackage{fix-cm}
\documentclass[twocolumn]{svjour3}          
\smartqed  
\usepackage{graphicx}

\usepackage{multirow}
\usepackage{amsfonts}
\usepackage{url}
\usepackage[switch]{lineno}
\usepackage{xcolor}
\usepackage{color}
\usepackage[textwidth=2.0cm,textsize=footnotesize]{todonotes}

\usepackage{fancyvrb}
\usepackage{balance}
\usepackage{booktabs}

\newcommand\mybold[1]{\textbf{\underline{#1}}}

\usepackage{comment}

\usepackage{ulem}
\newcommand{\aggiunta}[1]{\textcolor{black}{#1}}

\usepackage{amsmath}
\usepackage{amssymb}
\usepackage{hyperref}

\begin{document}

\title{\texttt{MINE GRAPH RULE}: A New GQL Operator for Mining Association Rules in Property Graph Databases}


\author{Francesco Cambria         \and
        Francesco Invernici         \and
        Anna Bernasconi         \and
        Stefano Ceri}

\institute{F. Cambria \at
              Via Ponzio 34/5, Politecnico di Milano, DEIB \\
              \email{francescoluciano.cambria@polimi.it}
           \and
           F. Invernici \at
              Via Ponzio 34/5, Politecnico di Milano, DEIB \\
              \email{francesco.invernici@polimi.it}
           \and
           A. Bernasconi \at
              Via Ponzio 34/5, Politecnico di Milano, DEIB \\
              \email{anna.bernasconi@polimi.it}
           \and
           S. Ceri \at
              Via Ponzio 34/5, Politecnico di Milano, DEIB \\
              \email{stefano.ceri@polimi.it}
}

\date{Received: 27 January 2025 / Accepted: 17 June 2025}

\maketitle

\begin{abstract}
Mining information from graph databases is becoming overly important. To approach this problem, current methods focus on identifying subgraphs with specific topologies; as of today, no work has been dedicated to jointly expressing the syntax and semantics of mining operations over rich property graphs.

We define \texttt{MINE GRAPH RULE}, a new operator for mining association rules from property graph databases, by following a research trend that has already been pursued for relational and XML databases. We describe the syntax and semantics of the operator, which allows measuring the support and confidence of each rule, and then 
we show many examples of increasing complexity, thereby providing a gentle introduction to the rich expressive power of the language, which is designed to be easy-to-use by GQL experts. 

Although the emphasis of this paper is on providing the syntax and semantics of the \texttt{MINE GRAPH RULE} operator, with several examples of use, we also developed an implementation of the 
operator on top of Neo4j, the most successful/adopted graph database system to date; the implementation is available as a portable Neo4j plugin, which we use to showcase real-world applications. 

At the end of our paper, we show the execution performance in a variety of synthetically generated settings, by varying the text of operators, the size of the graph, the ratio between node types, the method for creating relationships, and the maximum support and confidence; we also show our operator at work on two real-life graphs respectively describing music playlists and archived literature, and provide interesting examples of extracted association rules.
\end{abstract}


\section{Introduction}
Association rules for relational datasets take the form \(X \Rightarrow Y \), where the left-hand side \(X\) is denoted as body and the right-hand side \(Y\) is denoted as head
Both are sets instantiated from the same underlying domain (e.g., items), and their pairing within rules occurs statistically more often than ``normal'' in the context of certain groups of tuples within the same relational dataset (e.g., all tuples referring to the same transaction)~\cite{agrawal1993mining,agrawal1994fast,srikant1997mining}. Statistical significance is measured by two indexes, called support and confidence; the former is the probability of having both items \(X\) and \(Y\) within all transactions; the latter is the probability of having the \(Y\)  item in the transactions with the \(X\) item.

Association rules satisfy a classical antimonotonicity property stating that, for any two rules \(r_1: X_1 \Rightarrow Y \) and \(r_2: X_2 \Rightarrow Y \), with \(X_1 \supset X_2\) (e.g., \(X_1\) properly contains all the items of \(X_2\)), the support of \(r_1\)  is smaller or at most equal than the support of \(r_2\). The antimonotonicity property and the constraint on minimum support allow, when the support for a given association rule \(r: X \Rightarrow Y \) is below the threshold, to exclude also all rules whose body strictly includes \(X\)~\cite{vanetik2006support}. This property is at the basis of efficient mining rule methods, such as Apriori~\cite{apriori,santoso2021application}  or Equivalence Class Transformation Algorithm (ECLAT)~\cite{zaki2000scalable,kaur2015advanced}, which organize the search by arranging extracted bodies in a tree and cease exploring descendants of nodes that are not supported. 

The classical interpretation of association rules in recommender systems~\cite{changchien2001mining,lin2002efficient,osadchiy2019recommender} is to propose items that are often purchased in the same transaction, as an indication that they are jointly chosen by customers who -- in turn -- may share the same taste, although it is possible to have different interpretations of association rules for many other contexts of application, e.g., in the medical domain symptoms of patients affected by the same disease~\cite{stilou2001mining,nahar2013association,tandan2021discovering}, or in social networks interests that are shared by users~\cite{si2019association}.
Pattern languages have been designed for extracting given rules based upon application needs, e.g., focusing on given items in purchases (e.g., bread and butter), on given habits/morbidities in patients (e.g., male smokers), and on given interests among users (e.g., movies and music).

One such pattern language is the {\tt MINE RULE} operator, defined 
in~\cite{meo1996new}, which extracts semantically meaningful association rule patterns on relational datasets. The operator uses the expressive power of SQL and consists of several clauses, each mimicking an SQL query or predicate. Typically, {\tt MINE RULE} extracts tables with four columns 
[{\tt BODY}, {\tt HEAD}, {\tt SUPPORT}, {\tt CONFIDENCE}], each corresponding to an association rule, extracted from an underlying target table, constructed as an arbitrary SQL query expression whose tuples can be arbitrarily grouped and further partitioned; the table extracts all rules having
support and confidence beyond a given threshold. 

In the digital age, vast amounts of data are being generated and collected at an unprecedented pace~\cite{aggarwal2011introduction,parviainen2017tackling}. 
Relational databases are very effective in managing structured data, but face limitations when handling complex and interconnected data~\cite{he2008relational,moniruzzaman2013nosql}; graph databases are emerging as the leading data management technology for storing large knowledge graphs~\cite{vicknair2010comparison,robinson2015graph}. Therefore, 
there is a strong need to express association rules in this context. The main challenge when extending the broad work on relational association rule mining to graph-based association rule mining is to move from a simple, tabular data model, where associations are discovered within simple groups, to a complex graph-based data model, where associations can be semantically richer, and at the same time adapt the notions of support and confidence so as to preserve the antimonotonicity property, guaranteeing good convergence of the discovery methods. 

In this work, we present a new operator for graph databases, called \texttt{MINE GRAPH RULE}, which is inspired by the relational {\tt MINE RULE} operator~\cite{meo1996new}; the operator provides a declarative definition that takes advantage of the schema description available for property graphs (essentially labeled nodes and relationships with properties).  
We redefine the notion of `items' (introducing the novel concept of `anchor nodes') and progressively define the left and right sides of the association rules extracted by the {\tt MINE GRAPH RULE} operator; left and right sides are built by arbitrarily constructed orthogonal structures that support semantically rich path expressions, whose syntax and semantics are inspired by GQL, the emerging standard for graph query languages~\cite{deutsch2022graph,iso2024gql}. Our operator adapts the relational notion of {\it enclosing transactions} to property graphs, thereby allowing the computation of support and confidence for the pair of left and right sides extracted by the operator, along the classic semantics of association rules; as in \cite{meo1996new}, we extract association rules whose support and confidence is above given thresholds.

This work is different from most previous work in graph data mining, which typically focuses on searching for regular structures that form interesting sub-graphs that reflect given constraints upon their nodes or edges, without taking full advantage of the availability of schema description. Another main difference between our method and related work is that our declarative patterns can be translated to GQL-compliant Cypher and therefore be executed on graph database engines, taking advantage of their physical organization and query optimization methods, whereas most previous work is concerned with algorithmic approaches over ad-hoc data structures built from imported files, using programming languages such as Java~\cite{fan2015association} or C++~\cite{fan2022association}. 

\subsection*{Contributions} 

\noindent
This article presents five significant contributions:

\begin{itemize}
\item Syntax and semantics of the \texttt{MINE GRAPH RULE} operator. 
The operator supports the extraction of many items in the body and in the head, defined by a conjunction of relationship chains of GQL-compliant expressions of arbitrary length, which can be orthogonally composed. 
\item A rich set of progressive examples, which showcase the expressive power of the operator by extracting association rules of increasing complexity, both for what concerns the operator expressions and the complex tabular structure which returns the association rules whose support and confidence are above threshold.
\item Implementation and deployment of the operator to a real GQL-compliant~\cite{deutsch2022graph,iso2024gql} graph database. Among existing graph databases, ranked in \cite{dbranking}, we choose the top-ranked Neo4j system~\cite{jouili2013empirical,fernandes2018graph}.
We then describe how the \texttt{MINE GRAPH RULE} operator can be installed as a Neo4j plugin. The mining algorithm takes advantage of built-in optimizations of the Neo4j engine as well as optimizations that take advantage of the {\it Apriori} approach.
\item Performance evaluation of several data mining operators on artificially generated graphs, by varying several settings (including the artificial graph generation, graph size, relationship density, and minimum support/confidence).
\item Application of the operator to two large real-life datasets, respectively describing Spotify playlists and archived literature, and extraction of interesting association rules that showcase the potential of the approach.
\end{itemize}

\subsection*{Outline} 

This article is organized as follows: 
Section~\ref{sec:relwork} presents the related work;
Section~\ref{sec:syntax} describes the syntax and semantics of the \texttt{MINE GRAPH RULE} Operator 
and Section~\ref{sec:examples} illustrates several examples of increasing complexity, also showing the operator's output, produced as a table with several rows, one for each association rule, listing body, head, support, and confidence.
Section~\ref{sec:implementation} describes the implementation and the generation of the example graph database, and 
Section~\ref{sec:evaluation} describes the evaluation of our approach. Section \ref{sec:comparison} compares our work with PARM, a recently presented approach which also applies to property graphs.
Finally, Section~\ref{sec:conclusion} presents the discussion and conclusion.

\section{Related Work}
\label{sec:relwork} 

According to the panel discussion reported in~\cite{bonifati2025roadmap}, one of the key needs to improve graph database analytics is the development of more expressive languages, capable of supporting complex and diverse analyses. \texttt{MINE GRAPH RULE} aims to address this challenge, by providing a declarative pattern language that defines complex association rules for property graph databases.

\textbf{Declarative approaches for association rule mining.} 
In the existing data mining literature, several declarative pattern languages have been defined; among them, the highly cited MINE RULE operator (by Meo, Ceri, and Psaila)~\cite{meo1996new,meo1998extension} presents an SQL-based pattern language that extracts association rules calculated by using arbitrary data grouping and partitions; the operator produces readable outputs in tabular form, taking advantage of the NF2 model~\cite{NF2}, where each rule is associated with given support and confidence, enabling their filtering and ordering based on such statistical properties. The same authors provided a pipeline and framework to translate the MINE RULE operator to NF2 tables, using SQL~\cite{meo1998extension}. The MINE RULE approach and framework were then exploited by Boulicaut, Klemettinen, and Mannila, who define inductive databases as databases augmented by generic patterns and by an evaluation function telling how the pattern occurs in the data. In~\cite{boulicaut1998querying}, they explain that the MINE RULE approach perfectly fits the inductive database vision, being a {\it serious step towards an implementation framework for inductive databases}. The adaptation of MINE RULE to XML was proposed by Braga, Campi, Klemettinen, and Lanzi
~\cite{braga2002mining}, who defined XMINE as an operator to discover association rules based on the XQuery language.

\textbf{Graph pattern mining.} Most works in graph data mining look for regular structures in a graph, without taking advantage of a schema description; they search for interesting networks of nodes and edges, extracted from the graph, which reflect given constraints. For doing so, they typically read the graph into a memory structure and propose optimal algorithms, executed using programming languages such as Java or C++. Of course, the various definitions of {\it pattern} (GPAR, GFD, GAR, REE, PARM: see below) reflect the antimonotonicity property, which is the core of association rule mining. 

A framework for Graph Pattern Association Rule (GPAR) Mining is defined by Fan et al. in~\cite{fan2015association}, where 
a pattern is a subgraph is an arbitrary selection of nodes and edges from the graph (e.g., extracted by a query), and 
a rule is defined as follows: {\it when a pattern in which two designated nodes x and y are present is frequent in the graph, then also an edge between x and y is present};
for computing rule support, they use the minimum image-based support, as introduced in~\cite{bringmann2008frequent}, with a suitable revision to satisfy the antimonotonicity constraint. Their D-MINE algorithm uses ad-hoc auxiliary structures and takes advantage of parallelization to
extract rules that are very interesting (emerging from a top-k selection for diversified patterns).

This work was extended by Wang and Xu~\cite{wang2018mining}, who redefine GPAR, as rules are between two patterns such that each pattern is connected (includes connected nodes and edges) and two patterns cannot share any edge. The method is based on bisimulation as pattern matching semantics. They show that the problem can be decomposed into two steps, called frequent pattern mining and rule generation; their algorithm (FPMiner) exploits parallelism as well as look-ahead and backtracking to discover frequent patterns, defined as those above a minimal given threshold.  

Another interesting approach, also proposed by Fan et al., is concerned with mining graph functional dependencies (GFDs)~\cite{fan2016functional}, i.e. attribute-value dependencies and topological structures of entities, 
using an algorithm that explores trees progressively built out of graphs; the problem is coNP-complete, hence they develop an efficient parallel version that can cope with such complexity in specific contexts. Parallelism was later applied, by Fan et al.~\cite{fan2022association}, with application-driven reduction and sampling, to efficiently extract Graph Association Rules (GARs) from big graphs.

Recent work by Fan et al.~\cite{fan2022parallel} has discussed how to exploit parallelism and sampling in order to mine an extended set of Entity Enhancing Rules (REEs), which subsume functional dependencies and many other cases of dependencies -- named conditional dependencies, denial dependencies, and match dependencies (see~\cite{fan2022parallel}) suggested by Rock, an industrial system for data cleaning. Their method, called PRMiner, includes sampling and parallelism. Along this line of thought, recent work by Liu et al.~\cite{10.14778/3654621.3654641} has proposed a comprehensive framework designed to define and extract graph patterns by exploiting {\it oracles}, i.e., abstract machines developed for making decisions, typically either by importing external knowledge or by using internal computations such as aggregate operators or machine learning methods.

Other works address a broader problem, i.e. solving the so-called Frequent Subgraph Mining (FSM) problem \cite{new1}, defined as finding all the subgraphs within a graph that appear frequently (more than a give threshold); in general, the solution consists of two steps: generating candidate subgraphs and calculating their support, along with the Apriori method. \cite{new2} presents a method to prune options in the step generation, \cite{new3} presents a solution in a single pass.

More recently, Sasaki and Karras defined Path Association Rule Mining (PARM)~\cite{sasaki2024mining} as the problem of finding all the Path Association Rules (PAR) with specific thresholds on the paths' lengths and the rules' support. 
Given that PARM applies to property graphs and employs a similar approach to ours for the definition of graph association rules, it will be explored in greater detail in a dedicated section.

\textbf{Association rules mining using graphs managed by Neo4j.} 
Some works, with an application-oriented approach, address the extraction of association rules in real-world scenarios supported by Neo4j. Interesting case studies are reported in~\cite{ciftcci2021artist} (evaluating groups of artists appreciated by the same users) and in~\cite{standl2021pattern} (studying combinations of teaching methods that are more effective than others).
In both works, data is only extracted from Neo4j and structured in a way that the association rules are calculated thanks to established Python packages~\cite{apyori,raschkas_2018_mlxtend}. 
\cite{campi2021twitter} implements a case study in Neo4j to find popular Twitter hashtags that are posted together; the work includes an evaluation of the effect of pre-processing techniques that enhance the performance of the rule mining algorithm. In~\cite{sen2021recommendation}, a recommender system is fully developed in Neo4j. 
Note that all the mentioned works apply algorithms and the concepts of association rules to graph data, without formalizing rules and without fully exploiting the richness given by the natural structure of graphs.

\section{Operator Syntax and Semantics}
\label{sec:syntax}
\aggiunta{In graphs, the definition of association rules \(r: X \Rightarrow Y \) can be expanded considering $X$ and $Y$ not as unique data entries but as patterns of nodes and relationships. Our operator aims to define precisely and effectively these graph patterns.}

\begin{figure}[h!]
    \centering  
    \frame{\includegraphics[width=\linewidth]{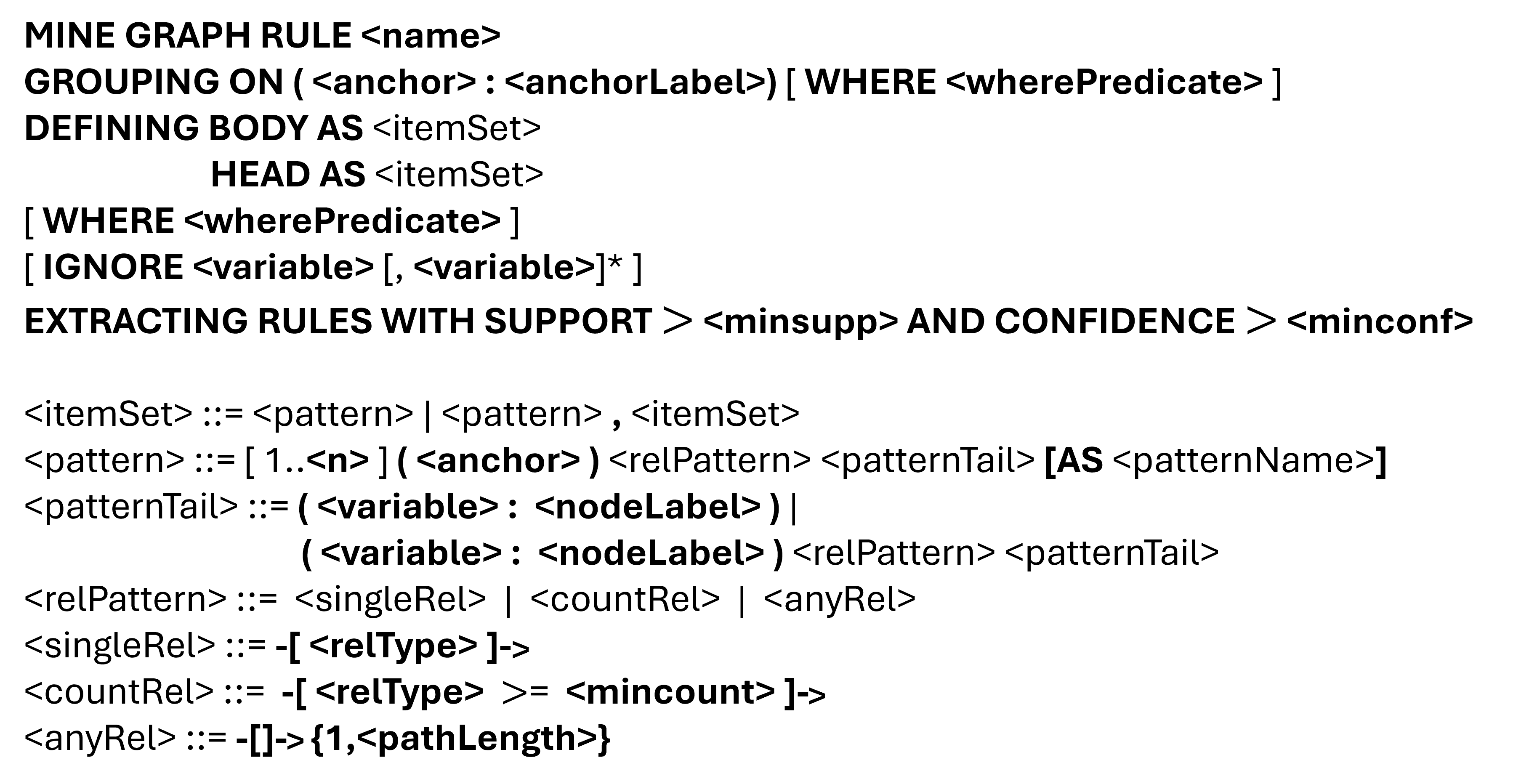}}
    \caption{\texttt{MINE GRAPH RULE} syntax. Uppercase bold type is used for terminal symbols.
    Lowercase is for non-terminal symbols; 
    bold symbols are not explained in the grammar, as their meaning is well-defined in the context of property graphs. 
    Note that square brackets are used to indicate optionality ([]) in the grammar;
    moreover, since \aggiunta{GQL} uses the symbols {\tt -[}  and {\tt ]-} for denoting relationships, square brackets are also used in bold-type to enclose the three expressions of \texttt{<relPattern>}. 
    Note as well that the $<$ and $>$ symbols, used to delimitate non-terminal symbols, are also used -in larger font size- to make comparisons in the pattern language: 
    $>$ is used to denote a `greater-than' comparison in the support/confidence threshold definition and in the \texttt{<countRel>} grammar production
    .}
    \label{fig:MINEGRAPHRULEOperatorSyntax}
\end{figure}

The syntax of \texttt{MINE GRAPH RULE} is shown in Fig.~\ref{fig:MINEGRAPHRULEOperatorSyntax}; 
it is inspired by the MINE RULE operator~\cite{meo1996new}, but it is significantly more complex to exploit the many ways of building associations by using property graph databases. It embeds some constructs of \aggiunta{GQL ~\cite{deutsch2022graph,iso2024gql}}
, so that it can easily be adapted for all the GQL-compatible database systems, and \aggiunta{it is} also more readable for graph database programmers. 

To each operator we associate a {\tt <name>}, allowing users to uniquely associate each operator with a set of extracted association rules.
Rules are built according to three expressions, respectively defining the {\tt GROUPING}, the rule's {\tt BODY}, and the rule's {\tt HEAD}. 
The grouping expression defines, through the \texttt{<anchorLabel>}, the set of {\it anchor nodes} of the graph that need to satisfy, when present, the {\tt WHERE} condition. They provide the {\it context of evaluation of association rules}; by analogy, the {\it anchor} acts as a pivot for the set of rules in the same way the purchase transaction groups the basket of items that are bought together. The {\tt BODY} and {\tt HEAD} expressions build respectively the left and right parts of an association rule; they are both defined by the \texttt{<itemSet>} production.

\sloppy{
An \texttt{itemSet} is defined as a conjunction of expressions, each called \texttt{<pattern>}, that extract the sets of items forming the head and body; the cardinality of each \texttt{<pattern>} varies between 1 (the default case) and a value {\tt <n>} (the cardinality must be defined only if {\tt <n>} is greater than 1).
A \texttt{<pattern>} generally consists of a linear path of relationships that starts from the anchor nodes (i.e., the nodes defined in the grouping expression) and can include an arbitrary number of relationship patterns (\texttt{<relPattern>});  thanks to an optional alias clause, introducing a user-defined {\tt <patternName>}, users can assign a custom name to each {\tt <pattern>}. 
Note that 
in patterns, every node needs a corresponding variable 
associated with a node label; using the same variable name is allowed only upon nodes with the same label, and implies the identity of the nodes extracted by the operator.

Recursion is introduced by the {\tt <patternTail>} production, which can either close the current recursion step with a final node or recursively include another {\tt <patternTail>}. 
Relationship patterns include three options (respectively denoted as {\tt <singleRel>}, {\tt <countRel>}, and {\tt <anyRel>}) whose explanation is postponed to three specific subsections. Note that -for simplicity- we assume that each relationship pattern is to be traversed from left to right, as explicitly expressed by the arrow direction.

The three expressions, {\tt GROUPING}, {\tt BODY}, and {\tt HEAD}, extract nodes that are referenced by variables; these can be used both to build any 
\aggiunta{GQL}-legal predicate 
\aggiunta{({\tt <wherePredicate>})} 
 for evaluating restricted parts of the graph database or for adding other grouping conditions defined by the optional {\tt IGNORE} clause with a list of at least one \texttt{<variable>}. 
The operator syntax is closed by setting two independent conditions for the {\it minimum support} {\tt <minsupp>} and {\it minimum confidence} {\tt <minconf>} that must be satisfied by the extracted rules.  

Concerning the semantics of pattern expressions, an important aspect is that the \texttt{MINE GRAPH RULE} uses the ``trail" path mode semantics~\cite{GQLstandard};
essentially, nodes can be revisited, but each relationship is used only once per path, to resolve the issue of cyclic instances. This is 
the most widely used semantics in graph database engines\footnote{In addition to ``trail", path modes include ``walk", ``acyclic", and ``simple". These path modes provide alternative path bindings, as discussed in Section 4.11.7 of the GQL standard~\cite{GQLstandard}.}.

Concerning the extraction of association rules, we need to calculate both support and confidence. We define $A$ as the set of anchor nodes; we denote $A$'s cardinality as $C(A)$. Evaluating the body and head expression matched with a set of anchor nodes $a \in A$ generates pairs $R_i= \langle B_i,H_i \rangle$, which are candidate association rules. Next we discuss how $B_i$ and $H_i$ are built.

Consider that, both in the BODY and HEAD of a rule, each {\tt <itemSet>} is associated with an ordered list of $m$ conjunctive expressions of {\tt <pattern>}s $P_m$, where each pattern corresponds to a list $L_m$ of named variables that are neither anchors nor included in the {\tt IGNORE} list of the pattern. Then, an operator evaluation, under trail semantics and different edges matching, maps each anchor node $a \in A$ (conceptually equivalent to a transaction) to a list $\mathbb{P}$ of lists $P_j$, with $1 \leq j \leq m$, such that each $P_j$ is a list of nodes of the graph obtained by matching the corresponding named variables $L_m$; a list $\mathbb{P}$ is separately computed for the body (producing $B_i$) and head (producing $H_i$). We then consider the mapping from the set $A$ of anchor nodes to the lists $B_i$ and $H_i$ and we denote as $C(B_i)$ the cardinality of the set of anchor nodes matched to $B_i$ and as $C(R_i)$ the cardinality of the set of anchor nodes matched to both $B_i$ and $H_i$. 


Then, for a given rule $R_i$, its support is $C(R_i)/C(A)$ (by analogy: the fraction of transactions having both the body and head items in the basket). Its confidence is $C(R_i)/C(B_i)$ (by analogy: the fraction of transactions that have both the body and the head over the transactions that have only the body). Finally, the operator extracts those candidate association rules that satisfy minimal constraints on support and confidence.

Each operator execution over a graph produces a non-normalized table (along the NF2 model~\cite{NF2}); the table is named as the operator and has four top-level columns. The first two columns carry names which are progressively constructed to reflect the specific pattern expressions for body and head (path expressions can be renamed using aliases);
the third and fourth columns contain support and confidence.

Each row of the table corresponds to an extracted association rule; the first and second columns respectively contain the complex structures $B_i$ and $H_i$, describing the rule's body and head.
Each node
forming the structures $B_i$ and $H_i$ 
is represented by system-generated node identifiers (\cite{GQLstandard}, Section 3.1.7); for ease of readability, for each node used in operators, we provide a map  ($<$NodeLabel$>$: $<$IdentifyingPropertyList$>$), where
the latter list is a user-provided identifier, made of suitable node properties, which are used instead of internal node identifiers for providing readable association rule instances\footnote{If a node does not have an identifier, it will appear in association rules as many times as it is extracted by the relevant pattern, without being able to give an identity to each node occurrence; but this could be considered as a poor combined design of the scheme and operator.}.

Note that, in the search for association rules, it is customary to exclude tautological rules, i.e., rules whose body and head are identical. Similarly, in the evaluation of the \texttt{MINE GRAPH RULE} pattern, we will exclude those rules whose body and head are identical. 


\section{Progressive Illustration of the Expressive Power of \texttt{MINE GRAPH RULE}}\label{sec:examples}
We next proceed to illustrate the expressive power of the operator, starting with a simple case that mimics the market basket analysis and then progressively addressing more complex cases.

\subsection{Running Example}

Fig.~\ref{fig:MotivatingExampleSchema} reports the schema of a property graph database, storing information about people's purchases that may reflect their social interactions; for easing interpretation, nodes and relationships of the running example are partitioned into subsets and each subset is associated with a specific label. The {\tt PERSON} nodes represent users of a social network, who can {\tt FOLLOW} each other and who can {\tt RECOMMEND} certain products represented by {\tt ITEM} nodes; the {\tt BUY} relationship indicates that given persons buy given products using an electronic marketplace, whereas the {\tt OF} relationship indicates a non-exclusive {\tt CATEGORY} which is assigned to each item by vendors on the basis of the expected context of use of that item. 

We also provide a super-small instance of the graph, shown in Fig.~\ref{fig:MotivatingExample}, in order to progressively introduce examples of \texttt{MINE GRAPH RULE} applications. Every node is uniquely identified by its {\tt Name} and carries specific properties: persons have {\tt Age} and  {\tt City}, items have {\tt Color} and {\tt Price}.  



\begin{figure}[ht]
    \centering    
    \includegraphics[width=.7\linewidth]{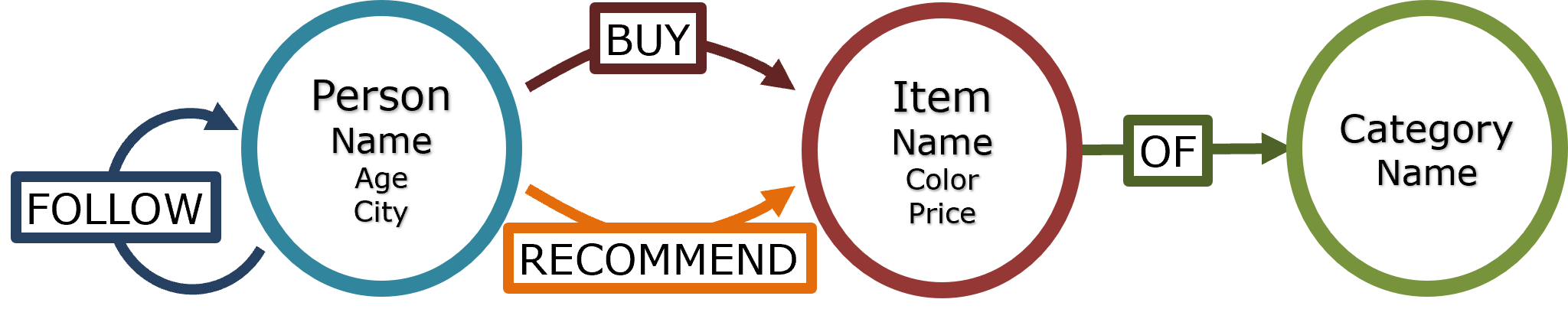}
    \caption{Schema of the running example}
    \label{fig:MotivatingExampleSchema}
\end{figure}

\begin{figure*}[h]
    \centering
    \includegraphics[width=0.9\linewidth]{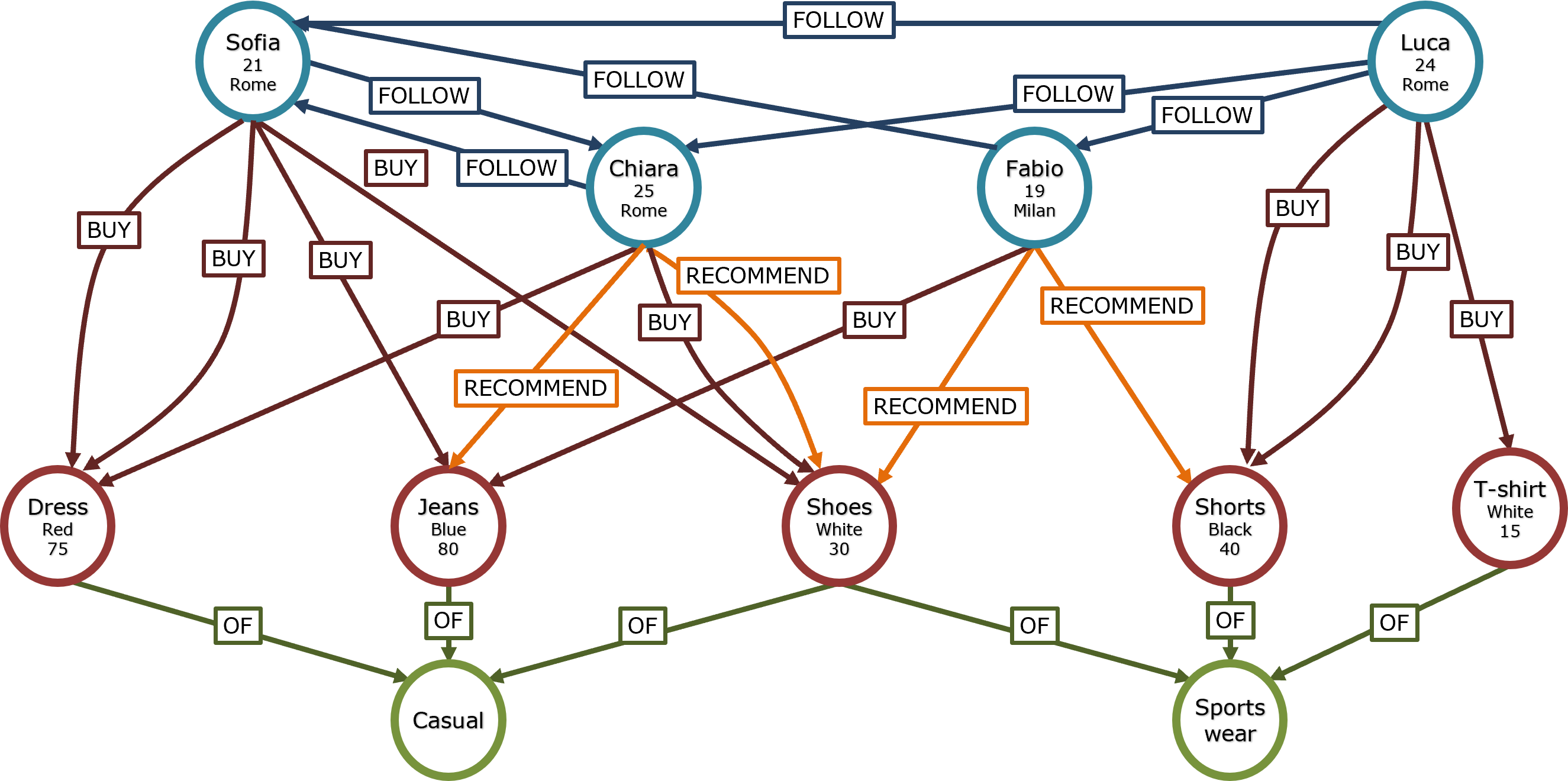}
    \caption{Instance of the running example}
    \label{fig:MotivatingExample}
\end{figure*}

\subsection{Simple Association Rules}
\label{sec:simpleassociationruleexample}

As a starting point, we show the most common application of association rules, i.e., pairs of different items usually bought together by the same person. These association rules are expressed by the natural language sentence: {\it ``People who buy X, also buy Y"}, where X and Y are different items. The \texttt{MINE GRAPH RULE} operator extracting these rules is simply: 

\begin{Verbatim}[frame=single, commandchars=\\\{\}, fontsize=\small]
MINE GRAPH RULE SimpleAssociationRules
GROUPING ON (p:Person)
DEFINING BODY AS (p)-[:BUY]->(X:Item)
         HEAD AS (p)-[:BUY]->(Y:Item)
EXTRACTING RULES WITH SUPPORT > 0.1
                  AND CONFIDENCE > 0.1         
\end{Verbatim}

The operator looks for all nodes labeled as {\tt (:Person)}. Both body and head expressions extract a single {\tt (:Item)}, connected by using the {\tt [:BUY]} relationship; hence the association rule combines the different items that are bought by the same person. Once applied to the graph database instance of Fig.~\ref{fig:MotivatingExample}, the operator produces the result shown in Table~\ref{tab:SimpleAssociationRuleTable}. The table's header has a hierarchical structure; the top of the hierarchy has four fixed attributes, respectively named {\tt Body}, {\tt Head}, {\tt Support}, and {\tt Confidence}. 
Below the {\tt Body} and {\tt Head}, the header includes terms describing the role of each extracted instance forming an association rule (in this case, {\tt BuyX} and {\tt BuyY}). The table's content includes one row for each association rule; {\tt Body} and {\tt Head} include the {\tt Name}s \aggiunta{(i.e., the identifying property of nodes)} describing the extracted Items.

\begin{table}[h!]
  \centering
  \resizebox{.7\columnwidth}{!}{
  \begin{tabular}{cccc}
    \toprule
    \textbf{Body} & \textbf{Head} & \multirow{2}{*}{\textbf{Support}} & \multirow{2}{*}{\textbf{Confidence}}\\ 
    \cmidrule(r){1-1}\cmidrule(l){2-2}
    \textbf{BuyX} & \textbf{BuyY} & &\\
    \midrule
    Dress & Jeans & 0.25 & 0.5\\ 
    Dress & Shoes & 0.5 & 1\\ 
    Jeans & Dress & 0.25 & 0.5\\ 
    Jeans & Shoes & 0.25 & 0.5\\ 
    Shoes & Dress & 0.5 & 1\\
    Shoes & Jeans & 0.25 & 0.5\\ 
    Shorts & T-shirt & 0.25 & 1\\ 
    T-shirt & Shorts & 0.25 & 1\\ 
    \bottomrule
  \end{tabular}
  }
  \caption{Output of SimpleAssociationRules}
  \label{tab:SimpleAssociationRuleTable}
\end{table}

Let us consider the first row of the table, corresponding to the rule: {\it ``People who buy Dress, also buy Jeans"}. 
For this row, $C(A)=4$, $C(B_1)=2$, $C(R_1)=1$. The first counter reports the total number of people (anchor node), the second the number of people who buy Dress (hence, Sofia and Chiara), and the third the total number of people who buy both Dress and Jeans (hence, just Sofia). Given these counters, the rule support is $C(R_1)/C(A)=0.25$, the rule confidence is $C(R_1)/C(B_1)=0.5$.

\subsection{Association Rules with Many Items in the Body or Head}
\label{sec:manyassociationruleexample}

Next, we show association rules with many items in the body or head. Consider the following \texttt{MINE GRAPH RULE} operator:

\begin{Verbatim}[frame=single, commandchars=\\\{\}, fontsize=\small]
MINE GRAPH RULE SimpleAssociationRulesWithManyItems
GROUPING ON (p:Person)
DEFINING BODY AS \mybold{1..2} (p)-[:BUY]->(X:Item)
         HEAD AS (p)-[:BUY]->(Y:Item)
EXTRACTING RULES WITH SUPPORT > 0.1
                  AND CONFIDENCE > 0.1         
\end{Verbatim}

The operator extracts the association rules whose body consists of up to two items, i.e., taking the form: {\it ``People who buy X$_1$ or X$_1$ and X$_2$, also buy Y"}; the body items are connected to the same anchor node person along the {\tt [:BUY]} relationship. The extracted association rules are reported in Table~\ref{tab:MoreItemsAssociationRulesTable}; note that the header now includes two columns for the body, respectively denoted as {\tt BuyX1} and {\tt BuyX2}. 
The rows of the table also include the rules extracted by the {\tt SimpleAssociationRules} operator (with a missing entry for the {\tt BuyX2} column).
 Three additional rows include two items in the body; among them, the rule of the first row indicates that {\it ``People who buy Dress and Jeans, also buy Shoes"}. 
Note that, for the first row, $C(A)=4$, $C(B_1)=1$, $C(R_1)=1$, hence the rule has support 0.25 and confidence 1. 

\begin{table}[h!]
  \resizebox{\columnwidth}{!}{
  \begin{tabular}{ccccc}
    \toprule
    \multicolumn{2}{c}{\textbf{Body}} & \textbf{Head} & \multirow{3}{*}{\textbf{Support}} & \multirow{3}{*}{\textbf{Confidence}}\\ 
    \cmidrule(r){1-2}\cmidrule(l){3-3}
    \textbf{ItemSetB1} & \textbf{ItemSetB2} & \textbf{ItemSetH1} & &\\ 
    \cmidrule(r){1-2}\cmidrule(l){3-3}
    \textbf{BuyX$_1$}  & \textbf{BuyX$_2$} & \textbf{BuyY} & &\\ \midrule
    Dress  & Jeans & Shoes & 0.25 & 1\\ 
    Dress  & Shoes & Jeans & 0.25 & 0.5\\ 
    Jeans  & Shoes & Dress & 0.25 & 1\\
    Dress  & - & Jeans & 0.25 & 0.5\\ 
    Dress  & - & Shoes & 0.25 & 1\\ 
    Jeans  & - & Dress & 0.25 & 0.5\\ 
    Jeans  & - & Shoes & 0.25 & 0.5\\ 
    Shoes  & - & Dress & 0.5 & 1\\ 
    Shoes  & - & Jeans & 0.25 & 0.5\\ 
    Shorts & - & T-shirt & 0.25 & 1\\ 
    T-shirt & - & Shorts & 0.25 & 1\\ 
    \bottomrule
  \end{tabular}
  }
  \caption{Output of SimpleAssociationRulesWithManyItems}
  \label{tab:MoreItemsAssociationRulesTable}
\end{table}


\subsection{Association Rules with Where Conditions}
\label{sec:whereassociationruleexample}

The evaluation of association rules can be restricted by adding simple predicates over the properties of variables defined in the \texttt{MINE GRAPH RULE} operator. 
For instance, it is possible to select the Person nodes by setting their {\tt City} to be ``Rome''. Adding a condition does not change the structure of the association rules, which are expressed as {\it ``People (living in Rome) who buy X, also buy Y''}.

\begin{samepage}
\begin{Verbatim}[frame=single, commandchars=\\\{\}, fontsize=\small]
MINE GRAPH RULE ConditionedAssociationRules1
GROUPING ON (p:Person) \mybold{WHERE p.city = "Rome"}
DEFINING BODY AS (p)-[:BUY]->(X:Item)
         HEAD AS (p)-[:BUY]->(Y:Item)
EXTRACTING RULES WITH SUPPORT > 0.1 
                  AND CONFIDENCE > 0.1      
\end{Verbatim}
\end{samepage}

Clearly, the restriction may reduce the cardinality of the anchor nodes, and thus produce different support and confidence for the extracted association rules. In our simple instance, the predicate is satisfied by Sofia, Chiara, and Fabio, hence $C(A)=3$. Table~\ref{tab:AliasConditionedAssociationRuleTable} reports the corresponding output.

\begin{table}[h!]
  \centering
  \resizebox{0.7\columnwidth}{!}{\begin{tabular}{cccc}
    \toprule
    \textbf{Body} & \textbf{Head} & \multirow{2}{*}{\textbf{Support}} & \multirow{2}{*}{\textbf{Confidence}}\\ 
    \cmidrule(r){1-1}\cmidrule(l){2-2}
    \textbf{BuyX} & \textbf{BuyY} & &\\ 
    \midrule
    Dress & Jeans & 0.33 & 0.5\\ 
    Dress & Shoes & 0.67 & 1\\ 
    Jeans & Dress & 0.33 & 1\\ 
    Shoes & Dress & 0.67 & 1\\ 
    Jeans & Shoes & 0.33 & 1\\ 
    Shoes & Jeans & 0.33 & 0.5\\ 
    Shorts & T-shirt & 0.33 & 1\\ 
    T-shirt & Shorts & 0.33 & 1\\ 
    \bottomrule
  \end{tabular}}
  \caption{Output of ConditionedAssociationRules1}
  \label{tab:AliasConditionedAssociationRuleTable}
\end{table}

Similarly, it is possible to define simple predicates on the properties of other variables of the body and head, for example by adding a restriction on the items' prices. 
The next example extracts the association rules: {\it ``People who buy X (with a price higher than 50), also buy Y (with a price lower than 50)''}: 

\begin{Verbatim}[frame=single, commandchars=\\\{\}, fontsize=\small]
MINE GRAPH RULE ConditionedAssociationRules2
GROUPING ON (p:Person)
DEFINING BODY AS (p)-[:BUY]->(X:Item) 
              AS HighPriceItem
         HEAD AS (p)-[:BUY]->(Y:Item) 
              AS LowPriceItem  
\mybold{WHERE X.Price > 50 and Y.Price < 50}       
EXTRACTING RULES WITH SUPPORT > 0.1 
                  AND CONFIDENCE > 0.1    
\end{Verbatim}

The resulting output is shown in Table~\ref{tab:BodyHeadConditionedAssociationRuleTable}. Note that, for better clarity, we renamed the output columns of the body and head.

\begin{table}[h!]
  \centering
  \resizebox{.7\columnwidth}{!}{
  \begin{tabular}{cccc}
    \toprule
    \textbf{Body} & \textbf{Head} & \multirow{2}{*}{\textbf{Support}} & \multirow{2}{*}{\textbf{Confidence}}\\ 
    \cmidrule(r){1-1}\cmidrule(l){2-2}
    \textbf{HighPriceItem} & \textbf{LowPriceItem} & &\\
    \midrule 
    Dress & Shoes & 0.5 & 1\\ 
    Jeans & Shoes & 0.25 & 0.5\\ 
    \bottomrule
  \end{tabular}
  }
  \caption{Output of ConditionedAssociationRules2}
  \label{tab:BodyHeadConditionedAssociationRuleTable}
\end{table}

\subsection{Association Rules with \texttt{CountRel}}
\label{sec:countassociationruleexample}

Next, we describe the use of the counting relations pattern, 
which is introduced by the {\tt <countRel>} non-terminal symbol in the grammar (Fig.~\ref{fig:MINEGRAPHRULEOperatorSyntax}). 
The pattern allows setting a threshold on the number of times a type of relationship should be present between any two nodes, in order to be selected in the
body or head of an association rule. 
The simplest example of application is to impose, in the body of the rule, that a given item is purchased several times. 
The operator should be used with a positive {\tt mincount} number to denote the threshold ({\tt <relType>} {\large $>=$} {\tt <mincount>}). An example application of the counting pattern is shown below:

\begin{Verbatim}[frame=single, commandchars=\\\{\}, fontsize=\small]
MINE GRAPH RULE CountItemsAssociationRules
GROUPING ON (p:Person)
DEFINING BODY AS (p)\mybold{-[:BUY>=2]->}(X:Item)
         HEAD AS (p)-[:BUY]->(Y:Item)
EXTRACTING RULES WITH SUPPORT > 0.1 
                  AND CONFIDENCE > 0.1      
\end{Verbatim}

The association rules are expressed in the sentence: {\it ``People who buy X at least 2 times, also buy Y''}. The corresponding output is shown in Table~\ref{tab:CountItemsAssociationRuleTable}.

\begin{table}[h!]
  \centering
  \resizebox{0.8\columnwidth}{!}{\begin{tabular}{cccc}
    \toprule
    \textbf{Body} & \textbf{Head} & \multirow{2}{*}{\textbf{Support}} & \multirow{2}{*}{\textbf{Confidence}}\\ 
    \cmidrule(r){1-1}\cmidrule(l){2-2}
    \textbf{BuyAtLeast2X} & \textbf{BuyY} & &\\
    \midrule
    Dress & Jeans & 0.25 & 1\\
    Dress & Shoes & 0.25 & 1\\
    Shorts & T-shirt & 0.25 & 1\\
    \bottomrule
  \end{tabular}}
  \caption{Output of CountItemsAssociationRules}
  \label{tab:CountItemsAssociationRuleTable}
\end{table}

Considering the association rule in the third row: {\it ``People who buy Shorts at least 2 times, also buy T-shirt''}, which associates people who buy the product Shorts two or more times with those who also buy T-shirt; Luca is the only person satisfying both conditions. The relevant counters for this rule are: $C(A)=4$, $C(B_3)=1$, $C(R_3)=1$, hence the rule has support 0.25 and confidence 1. 

\subsection{Association Rules with \texttt{AnyRel}}
\label{sec:anyassociationruleexample}

Next, we describe the use of the \texttt{AnyRel} pattern, which is introduced by the \texttt{<anyRel>} non-terminal symbol in the grammar (Fig.~\ref{fig:MINEGRAPHRULEOperatorSyntax}). The pattern allows us to connect a given node to any node reachable through any path (i.e., an arbitrary sequence of relationships) within a maximum length. The simplest example of application is to impose, in the body of the rule, that a given item is reached through at most one link (more complex examples are shown later). 
The operator should be used with a positive {\tt length} number to denote the path length. An example of the application is shown below:

\begin{Verbatim}[frame=single, commandchars=\\\{\}, fontsize=\small]
MINE GRAPH RULE AnyPathAssociationRules
GROUPING ON (p:Person)
DEFINING BODY AS (p)\mybold{-[]->\{1,1\}}(X:Item)
         HEAD AS (p)-[:BUY]->(Y:Item)
EXTRACTING RULES WITH SUPPORT > 0.1 
                  AND CONFIDENCE > 0.1      
\end{Verbatim}

Note that {\tt(:Item)} nodes are reached from the {\tt(:Person)} anchor nodes using either the {\tt [:BUY]} or the {\tt [:RECOMMEND]} relationships; thus, in our scenario, the sentence {\it ``People with any link to X, also buy Y''} can also be phrased as {\it ``People who buy or recommend X, also buy Y''}. The corresponding output is shown in Table~\ref{tab:AnyPathAssociationRulesTable}.

\begin{table}[h!]
  \centering
  \resizebox{0.7\columnwidth}{!}{\begin{tabular}{cccc}
    \toprule
    \textbf{Body} & \textbf{Head} & \multirow{2}{*}{\textbf{Support}} & \multirow{2}{*}{\textbf{Confidence}}\\ 
    \cmidrule(r){1-1}\cmidrule(l){2-2}
    \textbf{AnyLinkToX} & \textbf{BuyY} & &\\ \midrule
    Dress & Jeans & 0.25 & 0.5\\ 
    Dress & Shoes & 0.5 & 1\\ 
    Jeans & Dress & 0.5 & 0.67\\ 
    Jeans & Shoes & 0.5 & 0.67\\ 
    Shoes & Dress & 0.5 & 0.67\\ 
    Shoes & Jeans & 0.5 & 0.67\\ 
    Shorts & Jeans & 0.25 & 0.5\\ 
    Shorts & T-shirt & 0.25 & 0.5\\ 
    T-shit & Shorts & 0.25 & 1\\ \bottomrule
  \end{tabular}}
  \caption{Output of AnyPathAssociationRules}
  \label{tab:AnyPathAssociationRulesTable}
\end{table}

In order to better understand it, consider the fourth association rule of the table, i.e., {\it ``People who buy or recommend Jeans, also buy Shoes''}.
As usual, $C(A)=4$. Sofia, Chiara, and Fabio buy or recommend Jeans, thus $C(B_4)=3$; out of them, just Sofia and Chiara buy Shoes, thus $C(R_4)=2$.  
Given these counters, the rule support is $C(R_4)/C(A)=0.5$, the rule confidence is $C(R_4)/C(B_4)=0.67$.

By increasing the {\tt length} parameter, the operator evaluates longer patterns, introducing more alternatives. For instance, consider the following operator:

\begin{Verbatim}[frame=single, commandchars=\\\{\}, fontsize=\small]
\textcolor{black}{MINE GRAPH RULE}
\textcolor{black}{AnyLongerPathAssociationRules}
\textcolor{black}{GROUPING ON (p:Person)}
\textcolor{black}{DEFINING BODY AS (p)\mybold{-[]->\{1,2\}}(X:Item)}
\textcolor{black}{         HEAD AS (p)-[:BUY]->(Y:Item)}
\textcolor{black}{EXTRACTING RULES WITH SUPPORT > 0.1 }
\textcolor{black}{                  AND CONFIDENCE > 0.1}      
\end{Verbatim}

The operator extracts the association rules taking form: {\it ``People with any link (of length up to 2) to X, also buy Y''}. In this scenario, nodes {\tt (:Person)}
can reach nodes {\tt (:Item)} either directly, by buying or recommending them (as the previous example), or also with two-hops paths, by using the relationship {\tt [:FOLLOW]} connecting a node {\tt (:Person)} to another node {\tt (:Person)}, then connected to the item with a {\tt [:BUY]} or {\tt [:RECOMMEND]} relationship. In Table~\ref{tab:AnyPathLongerAssociationRulesTable}, the resulting association rules are reported.

\begin{table}[h!]
  \centering
  \resizebox{0.9\columnwidth}{!}{\begin{tabular}{cccc}
    \toprule
    \textbf{Body} & \textbf{Head} & \multirow{2}{*}{\textbf{Support}} & \multirow{2}{*}{\textbf{Confidence}}\\ 
    \cmidrule(r){1-1}\cmidrule(l){2-2}
    \textbf{AnyLinkOfLengthUpTo2ToX} & \textbf{BuyY} & &\\ \midrule
    Dress & Jeans & 0.5 & 0.5\\ 
    Dress & Shoes & 0.5 & 0.5\\ 
    Dress & Shorts & 0.25 & 0.25\\
    Dress & T-shirt & 0.25 & 0.25\\ 
    Jeans & Dress & 0.5 & 0.5\\ 
    Jeans & Shoes & 0.25 & 0.25\\ 
    Jeans & Shorts & 0.25 & 0.25\\ 
    Jeans & T-shirt & 0.25 & 0.25\\ 
    Shoes & Dress & 0.5 & 0.5\\ 
    Shoes & Jeans & 0.5 & 0.5\\ 
    Shoes & Shorts & 0.25 & 0.25\\ 
    Shoes & T-shirt & 0.25 & 0.25\\ 
    Shorts & Jeans & 0.25 & 0.25\\ 
    Shorts & T-shirt & 0.25 & 0.5\\ 
    T-shit & Shorts & 0.25 & 1\\ \bottomrule
  \end{tabular}}
  \caption{Output of AnyLongerPathAssociationRules}
  \label{tab:AnyPathLongerAssociationRulesTable}
\end{table}

Consider the seventh association rule {\it ``People with any link (of length up to 2) to Jeans also buy Shorts"}, which is not included among the extracted rules in the previous example. We still have $C(A) = 4$, but while Sofia, Chiara, and Fabio directly buy or recommend Jeans, Luca follows Sofia, who in turns buys Jeans; therefore, $C(B_7) = 4$ instead of $3$. Among them, only Luca buys Shorts, so $C(R_7) = 1$. The resulting support of the rule is $C(R_7)/C(A) = 0.25$, and the confidence of the rule is $C(R_7)/C(B_7) = 0.25$.

\subsection{Association Rules with Relationships' Chains}
\label{sec:chainassociationruleexample}

Next, we consider that {\tt <relPattern>}s can be chained, in the body, in the head, or in both of them; as the grammar has a recursive production
(\texttt{<patternTail>}), the length of the chain is not bound. 

We consider a path in the body of the \texttt{MINE GRAPH RULE} operator, linking purchased items to their categories, so as to include also categories in resulting association rules. Consider the following operator:

\begin{Verbatim}[frame=single, commandchars=\\\{\}, fontsize=\small]
MINE GRAPH RULE PathAssociationRules1
GROUPING ON (p:Person)
DEFINING BODY AS (p)-[:BUY]->(X:Item)
                             \mybold{-[:OF]->(C:Category)}
         HEAD AS (p)-[:BUY]->(Y:Item)
EXTRACTING RULES WITH SUPPORT > 0.1 
                  AND CONFIDENCE > 0.1 
\end{Verbatim}

Note that the {\tt(:Category)} nodes are reached from the {\tt(:Person)} anchor nodes along a chain of relationships, which uses the {\tt [:BUY]} and {\tt [:OF]} relationships; the interpretation of the generated association rules is {\it ``People who buy X of category C, also buy Y''}. The corresponding output is shown in Table~
\ref{tab:LongerItemsAssociationRulesTable}. 

\begin{table}[h!]
\centering
  \resizebox{.85\columnwidth}{!}{\begin{tabular}{ccccc}
    \toprule
    \multicolumn{2}{c}{\textbf{Body}} & \textbf{Head} & \multirow{3}{*}{\textbf{Support}} & \multirow{3}{*}{\textbf{Confidence}}\\ \cmidrule(r){1-2}\cmidrule{3-3}
    \multicolumn{2}{c}{\textbf{ItemSetB1}} & \textbf{ItemSetH1} & &\\ \cmidrule(r){1-2}\cmidrule{3-3}
    \textbf{BuyX}  & \textbf{OfC} & \textbf{BuyY} & &\\ \midrule
    Dress  & Casual & Jeans & 0.25 & 0.5\\ 
    Dress  & Casual & Shoes & 0.5 & 1\\ 
    Jeans  & Casual & Dress & 0.25 & 0.5\\ 
    Shoes  & Casual & Dress & 0.5 & 1\\ 
    Shoes  & Sportswear & Dress & 0.5 & 1\\ 
    Shoes  & Casual & Jeans & 0.25 & 0.5\\ 
    Shoes  & Sportswear & Jeans & 0.25 & 0.5\\ 
    Shorts & Sportswear & T-shirt & 0.25 & 1\\ 
    T-shirt & Sportswear & Shorts & 0.25 & 1\\ \bottomrule
  \end{tabular}}
  \caption{Output of PathAssociationRules1}
  \label{tab:LongerItemsAssociationRulesTable}
\end{table}

If we consider the first association rule, {\it ``People who buy Dress of category Casual also buy Jeans''}, we note that Sofia and Chiara buy Dress of category Casual but just Sofia buys Jeans; hence, with $C(A)=4$, we have $C(B_1)=2$ and $C(R_1)=1$. Given these the counters, the rule support is $C(R_1)/C(A)=0.25$, the rule confidence is $C(R_1)/C(B_1)=0.5$.

\subsection{Association Rules with Ignore}
\label{sec:ignoreassociationruleexample}

The {\tt IGNORE} construct designates a list of variables -i.e., nodes of the pattern- that shall not be included in the body or head of the rule. We denote as {\it visible variables} those appearing in the operator but not in the ignore clause; only visible variables will appear in the body and head of association rules. 

For example, consider the following \texttt{MINE GRAPH RULE} operator: 

\begin{Verbatim}[frame=single, commandchars=\\\{\}, fontsize=\small]
MINE GRAPH RULE IgnoreAssociationRules
GROUPING ON (p:Person)
DEFINING BODY AS (p)-[:BUY]->(X:Item)
                             -[:OF]->(CX:Category)
         HEAD AS (p)-[:BUY]->(Y:Item)
                             -[:OF]->(CY:Category)
\mybold{IGNORE X,Y}
EXTRACTING RULES WITH SUPPORT > 0.1 
                  AND CONFIDENCE > 0.1
\end{Verbatim}

Extracted rules take the form {\it ``People who buy items of category CX, also buy items of category CY''}. 
The corresponding output is shown in Table~\ref{tab:IgnoreAssociationRules}. 

\begin{table}[h!]
  \resizebox{1\columnwidth}{!}{\begin{tabular}{cccc}
    \toprule
    \textbf{Body} & \textbf{Head} & \multirow{3}{*}{\textbf{Support}} & \multirow{3}{*}{\textbf{Confidence}}\\ 
    \cmidrule(r){1-1}\cmidrule(l){2-2}
    \textbf{ItemSetB1} & \textbf{ItemSetH1} & &\\ 
    \cmidrule(r){1-1}\cmidrule(l){2-2}
    \textbf{BuyItemsOfCategoryCX} & \textbf{BuyItemsOfCategoryCY} & &\\ 
    \midrule
    Casual & Sportswear & 0.5 & 0.67\\ 
    Sportswear & Casual & 0.5 & 0.67\\ \bottomrule
  \end{tabular}}
  \caption{Output of IgnoreAssociationRules}
  \label{tab:IgnoreAssociationRules}
\end{table}

 
Using {\tt IGNORE} adds a second aggregation level to the normal procedure: in particular, after having extracted the anchor nodes for body and head, they are further grouped only with respect to the {\it visible variables}, thus hiding the information of the {\it ignored variables}. 
Consider rule 1, {\it ``People who buy items of category Casual, also buy items of category Sportswear''}. 
Sofia, Chiara, and Fabio buy at least one item in the category Casual, resulting in $C(B_1)=3$, but Sofia and Chiara also buy items in the category Sportwear, whereas Fabio does not buy any item in that category. As Fabio is not paired to both the body and the head, Fabio is not a valid anchor for the association rule, and $C(R_1)=2$. 
As usual, with $C(A)=4$ the rule support is $C(R_1)/C(A)=0.5$, the rule confidence is $C(R_1)/C(B_1)=0.67$.

\subsection{Association Rules with Conjunctions}
\label{sec:and1associationruleexample}
Next, we consider that both the body and the head can have conjunctive expressions, with an arbitrary number of conjuncts.
We consider the body of the \texttt{MINE GRAPH RULE} operator as a conjunction of two different {\tt <pattern>}s. Consider the following operator:

\begin{Verbatim}[frame=single, commandchars=\\\{\}, fontsize=\small]
MINE GRAPH RULE ComplexBodyAssociationRules
GROUPING ON (p:Person)
DEFINING BODY AS (p)-[:FOLLOW]->(X:Person)\mybold{,}
                            \mybold{(p)-[:BUY]->(Y:Item)}
         HEAD AS (p)-[:BUY]->(Z:Item)
EXTRACTING RULES WITH SUPPORT >  0.1 
                  AND CONFIDENCE > 0.1      
\end{Verbatim}

The association rules are interpreted in this way: {\it ``People who follow X and buy Y, also buy Z''}. The corresponding output is shown in Table~\ref{tab:ComplexBodyAssociationRulesTable}. Considering the first association rule, {\it ``People who follow Sofia and buy Dress, also buy Shoes''}, we note that Chiara is the only one who both follows Sofia and buys Dress, and she also buys Shoes; 
hence, with $C(A)=4$, we have $C(B_1)=1$ and $C(R_1)=1$ and with these counters, the rule support is $C(R_1)/C(A)=0.25$ and the rule confidence is $C(R_1)/C(B_1)=1$.

\begin{table}[h!]
  \resizebox{\columnwidth}{!}{\begin{tabular}{ccccc}
    \toprule
    \multicolumn{2}{c}{\textbf{Body}} & \textbf{Head} & \multirow{3}{*}{\textbf{Support}} & \multirow{3}{*}{\textbf{Confidence}}\\ 
    \cmidrule(r){1-2}\cmidrule(l){3-3}
    \textbf{ItemSetB1} & \textbf{ItemSetB2} & \textbf{ItemSetH1} & &\\ 
    \cmidrule(r){1-2}\cmidrule(l){3-3}
    \textbf{FollowX}  & \textbf{BuyY} & \textbf{BuyZ} & &\\ \midrule
    Sofia  & Dress & Shoes & 0.25 & 1\\ 
    Sofia  & Shoes & Dress & 0.25 & 1\\ 
    Sofia  & Shorts & T-shirt & 0.25 & 1\\ 
    Sofia  & T-shirt & Shorts & 0.25 & 1\\ 
    Chiara & Dress & Jeans & 0.25 & 1\\ 
    Chiara & Dress & Shoes & 0.25 & 1\\ 
    Chiara & Jeans & Dress & 0.25 & 1\\ 
    Chiara & Jeans & Shoes & 0.25 & 1\\ 
    Chiara & Shoes & Dress & 0.25 & 1\\ 
    Chiara & Shoes & Jeans & 0.25 & 1\\ 
    Chiara & Shorts & T-shirt & 0.25 & 1\\ 
    Chiara & T-shirt & Shorts & 0.25 & 1\\ 
    Fabio & Shorts & T-shirt & 0.25 & 1\\ 
    Fabio & T-shirt & Shorts & 0.25 & 1\\ \bottomrule
    \end{tabular}}
  \caption{Output of ComplexBodyAssociationRules}
  \label{tab:ComplexBodyAssociationRulesTable}
\end{table}

\subsection{Other Complex Association Rules}

\subsubsection{First Example.}
We consider a body {\tt <pattern>}, with two conjuncts both consisting of chains of relationships. The following operator extracts rules whose interpretation is {\it ``People who follow more than two people who recommend products of category CX and buy products of category CY, also buy Z''}.
The corresponding output is shown in Table~\ref{tab:MoreComplexBodyAssociationRulesTable}. Considering the first association rule, ``People who follow more than two persons who recommend an item of category Sportswear and also buy items of category Sportswear, also buy T-shirt''; we note that Luca is the only one who both follows more than two persons -Chiara and Fabio- recommending Shoes and Shorts of category Sportswear, and also buys items of category Sportswear; therefore, with $C(A)=4$, we have $C(B_1)=1$ and $C(R_1)=1$ and, with these counters, the rule support is $C(R_1)/C(A)=0.25$ and the rule confidence is $C(R_1)/C(B_1)=1$.

\begin{Verbatim}[frame=single, commandchars=\\\{\}, fontsize=\small]
MINE GRAPH RULE MoreComplexBodyAssociationRules
GROUPING ON (p:Person)
DEFINING BODY AS (p)-[:FOLLOW>=3]->(PX:Person)
                            -[:RECOMMEND]->(X:Item)
                            -[:OF]->(CX:Category),
                 (p)-[:BUY]->(Y:Item)
                            -[:OF]->(CY:Category)
         HEAD AS (p)-[:BUY]->(Z:Item)
IGNORE PX, X, Y
EXTRACTING RULES WITH SUPPORT > 0.1 
                  AND CONFIDENCE > 0.1      
\end{Verbatim}

\begin{table}[h!]
  \resizebox{\columnwidth}{!}{\begin{tabular}{ccccc}
    \toprule
    \multicolumn{2}{c}{\textbf{Body}} & \textbf{Head} & \multirow{3}{*}{\textbf{Support}} & \multirow{3}{*}{\textbf{Confidence}}\\ 
    \cmidrule(r){1-2}\cmidrule(l){3-3}
    \textbf{ItemSetB1} & \textbf{ItemsSetB2} & \textbf{ItemSetH1} & &\\ 
    \cmidrule(r){1-2}\cmidrule(l){3-3}
    \textbf{FollowPersonRecommendItemOfCX}  & \textbf{BuyItemOfCY} & \textbf{BuyZ} & &\\ \midrule
    Sportswear & Sportswear & T-shirt & 0.25 & 1 \\ 
    Sportswear & Sportswear & Shorts & 0.25 & 1 \\ 
    Casual & Sportswear & T-shirt & 0.25 & 1 \\
    Casual & Sportswear & Shorts & 0.25 & 1 \\ \bottomrule
    \end{tabular}}
  \caption{Output of MoreComplexBodyAssociationRules }
  \label{tab:MoreComplexBodyAssociationRulesTable}
\end{table}

\subsubsection{Second Example.} 
Lastly, we consider a complex structure of the body {\tt <pattern>} with {\tt WHERE} conditions on both anchors and nodes. The following operator extracts rules whose interpretation is {\it ``People from Rome who follow Chiara, who recommends items of category CX, also buy items of the same category''}.
The corresponding output is shown in Table~\ref{tab:VerifyInfluencerEffectivenessInRome}.
Considering the first association rule, \textit{``People from Rome who follow Chiara who recommends items of category Casual, also buy Jeans, which are of the same category Casual''}, we note that there is only one person, Sofia, who satisfies all these constraints, but also Luca follows Chiara recommending Casual items; hence, with $C(A)=3$, we have $C(B_1)=2$ and $C(R_1)=1$ and, with these counters, the rule support is $C(R_1)/C(A)=0.33$ and the rule confidence is $C(R_1)/C(B_1)=0.5$.


\begin{Verbatim}[frame=single, commandchars=\\\{\}, fontsize=\small]
MINE GRAPH RULE VerifyInfluencerEffectivenessInRome
GROUPING ON (p:Person) WHERE p.city = "Rome"
DEFINING BODY AS (p)-[:FOLLOW]->(PX:Person)
                            -[:RECOMMEND]->(X:Item)
                            -[:OF]->(CX:Category)
         HEAD AS (p)-[:BUY]->(Y:Item)
                            -[:OF]->(CY:Category)
WHERE PX.name = "Chiara" AND CX.name = CY.name
IGNORE PX, X
EXTRACTING RULES WITH SUPPORT > 0.1 
                  AND CONFIDENCE > 0.1 
\end{Verbatim}

\begin{table}[h!]
  \resizebox{\columnwidth}{!}{\begin{tabular}{ccccc}
    \toprule
    \textbf{Body} & \multicolumn{2}{c}{\textbf{Head}} & \multirow{3}{*}{\textbf{Support}} & \multirow{3}{*}{\textbf{Confidence}}\\ 
    \cmidrule(r){1-1}\cmidrule(l){2-3}
    \textbf{ItemSetB1} & \multicolumn{2}{c}{\textbf{ItemSetH1}} & &\\ 
    \cmidrule(r){1-1}\cmidrule(l){2-3}
    \textbf{FollowChiaraRecommendCX}  & \textbf{BuyItem} & \textbf{OfCY} & &\\ \midrule
    Casual & Jeans & Casual & 0.33 & 0.5 \\ 
    Casual & Dress & Casual & 0.33 & 0.5 \\ 
    Casual & Shoes  & Casual & 0.33 & 0.5 \\ 
    Sportswear & Shoes  & Sportswear & 0.33 & 0.5 \\ 
    Sportswear & T-shirt  & Sportswear & 0.33 & 0.5 \\ 
    Sportswear & Shorts & Sportswear & 0.33 & 0.5 \\ \bottomrule 
    \end{tabular}}
  \caption{Output of VerifyInfluencerEffectivenessInRome }
  \label{tab:VerifyInfluencerEffectivenessInRome}
\end{table}

\begin{figure*}[!ht]
    \centering
    \includegraphics[trim={0 10cm 0 0},clip,width=\linewidth]{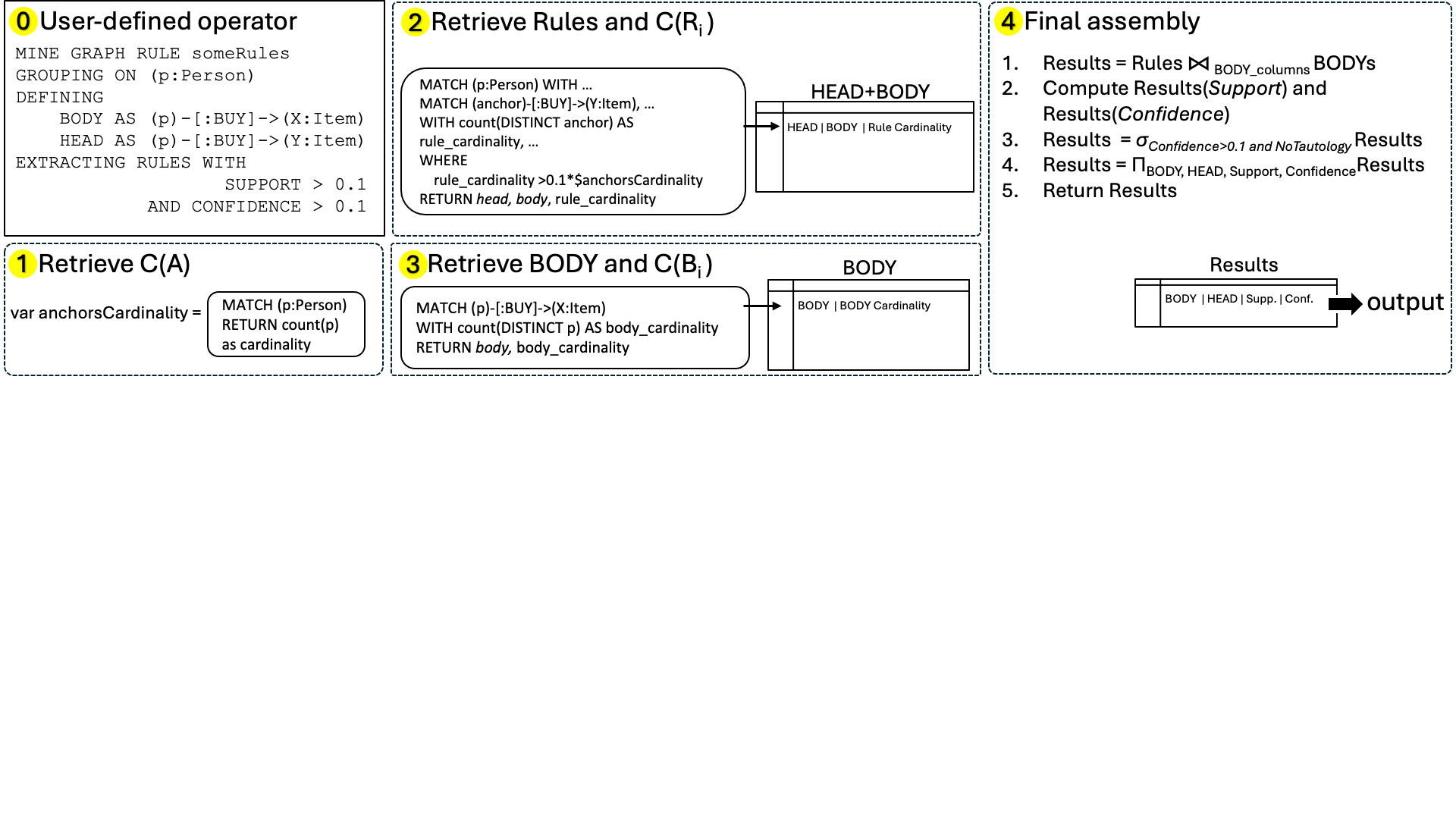}
    \caption{Phases of the \texttt{MINE GRAPH RULE} algorithm applied to a simple user-defined operator. 
{\tt Phase(1)} counts the anchor nodes with the label `Person' $C(A)$. 
In {\tt Phase(2)}, a query generating head and body for the operator is issued, and the query results, which satisfy the minimum support threshold (expressed using absolute counts), are entered into a table whose entries include the HEAD, the BODY, and the rule 
cardinality $C(R_i)$
In {\tt Phase(3)}, a query generating just the body 
for the operator is issued and the query results are entered into a table whose entries include the BODY and the BODY
cardinality $C(B_i)$.
In the final {\tt Phase(4)} the output table is produced along a progression of five steps: 
first, the two tables produced by phases 3 and 4 are joined on the BODY columns; 
then, support and confidence for each record of the resulting table are computed; 
then, rows that create tautologies or do not satisfy the minimum confidence threshold are removed; 
then, the result is projected upon the chosen schema;
finally, resulting rows are streamed in output.
}
    \label{fig:algo_base}
\end{figure*}

\section{Implementation}
\label{sec:implementation}
\sloppy{
Since the GQL standard was released only in April 2024, mature products that fully adopt the standard are in progress. We chose to base our implementation on Cypher~\cite{francis2018cypher}, the query language of Neo4j, due to its close affinity to GQL.
We selected the Neo4j Graph database as a development platform since it provides an easily expandable architecture. Still, our proposed implementation does not strictly depend on any specific Neo4j component or Cypher construct.
Implementations for other graph databases and frameworks, like Memgraph, JanusGraph, or Apache TinkerPop, can be developed with no particular efforts. 

Our implementation of the \texttt{MINE GRAPH RULE} operator has been developed in Java as a user-defined procedure; it is available as a Neo4j plugin like other add-ons of the Awesome Procedures on Cypher (APOC)~\cite{apoc} library.  Libraries providing the data structures exploited in the current implementation are also available for most of the other main graph database systems. 
}

Once the plugin is installed, the procedure can be invoked by its name ``{\tt mineGraphRule}'' in the Neo4j Cypher shell; 
it supports nine mandatory parameters that map one-to-one to the variable elements of the operator contained in the tuple $(a, A, W_a, H, B, W, I, s, c)$:
the anchor's variable $a$, 
the anchor's label $A$, 
the conditions $W_a$ on the anchor, 
the map structure $H$ for the \texttt{HEAD} itemSet, 
the map structure $B$ for the \texttt{BODY} itemSet, 
the list  of predicates $W$ for the \texttt{WHERE} clause, 
the list of \texttt{IGNORE}d variables $I$, 
the minimum support threshold $s$, and 
the minimum confidence threshold $c$.
A detailed specification of the parameters' format can be found in the project's repository~\cite{cambria2024repo-ORIGINAL}.
Collectively, these nine parameters provide a semi-structured version of the \texttt{MINE GRAPH RULE} operator. 
In our implementation, we associated each node with a system-generated identifier, to ensure the uniqueness of each node.


Once executed, the procedure outputs a stream of records (each representing one association rule) with a fixed format, consisting of four fields:
\begin{itemize}
    \item {\tt body}: a map of key-value pairs for the {\tt BODY} columns. The keys are the column names
    , whereas the values are the identifiers (i.e., properties) of the nodes 
    matched by the specified patterns;
    \item {\tt head}: a map of key-value pairs for the {\tt HEAD} columns, with the same format of the {\tt BODY};
    \item {\tt support}: the support value in double floating-point precision;
    \item {\tt confidence}: the confidence value in double floating-point precision.
\end{itemize}
 
By default, the APOC procedure uses each internal node identifier to match the nodes and generate the output table for both the {\tt BODY} and {\tt HEAD} columns. However, with an optional input parameter, the user can provide a list of identifying attributes as key-value pairs ($<$NodeLabel$>$: $<$IdentifyingPropertyList$>$) specifying, for each or some node label, which property should be used instead of the internal node identifier.

The result can be transformed with the Cypher \texttt{YIELD} sub-clause~\cite{cypherYield}, which can 
select any column of the output record and apply any supported Cypher or APOC operation on them, such as reordering, regrouping, or string concatenation.

We next describe our algorithm by first considering a base case, not including conjunctions or multiple items in the BODY and in the HEAD. Then, we consider the general case.

\subsection{Algorithm for Simple Cases}
\label{sec:algo_simple}

In the translation, we use a syntax-directed approach in which, for every production of the grammar, we generate an appropriate Cypher expression. 
Therefore, 
chains are directly translated as Cypher {\tt MATCH} blocks targeting alternatively: a simple relationship, a relationship filtered based on the number of edges, or a bounded path with free-type relationships.
In this way, our \texttt{MINE GRAPH RULE} language, which includes a rich collection of orthogonal features, can be expressed in a compact Cypher query, whose execution takes advantage of the optimization capabilities of graph database engines.
In this simple case, we only need the following queries:
\begin{itemize}
    \item a query for retrieving the anchors' cardinality $C(A)$ (see Section~\ref{sec:syntax});
    \item a query for retrieving HEAD, BODY, and rule's cardinality $C(R_i)$. Since this is the most selective query, we obtain the rules' confidence at an early stage, so that an optimized execution can discard all rules below the confidence threshold;
    \item for the remaining rules, a query for retrieving BODY and body's cardinality $C(B_i)$. At this point, the rules' support can be computed and the optimized query execution can discard all rules below the support threshold.
\end{itemize}
The first query follows a simple structure and produces a variable counting the number of anchors.
The other two queries have a similar structure, which is described as follows:
\begin{enumerate}
    \item {\tt MATCH} subquery for anchors' selection based on labels and on predicates that are specific to anchors' properties;
    \item {\tt MATCH} on all BODY (and possibly HEAD) patterns in conjunction;
    \item optional {\tt WHERE} predicates, when operators include conditions on the variables;
    \item {\tt WITH} clause to group by anchors and visible variables, to compute the rule cardinality;
    \item {\tt WHERE} predicate to guarantee minimum support;
    \item {\tt RETURN} clause with the final names returned to the user at the end of the procedure.  
\end{enumerate}
Cypher queries produced for bodies only include BODY's patterns, whereas queries produced for rules also include elements from heads.

The actual association rules, with their support and confidence, are produced after a final assembly of the results of the three described Cypher queries.
Fig.~\ref{fig:algo_base} shows the complete flow of execution of the graph association rule mining algorithm when applied to a simple operator. 

\begin{figure*}[!ht]
    \centering
    \includegraphics[trim={0 0cm 0cm 0},clip,width=.72\linewidth]{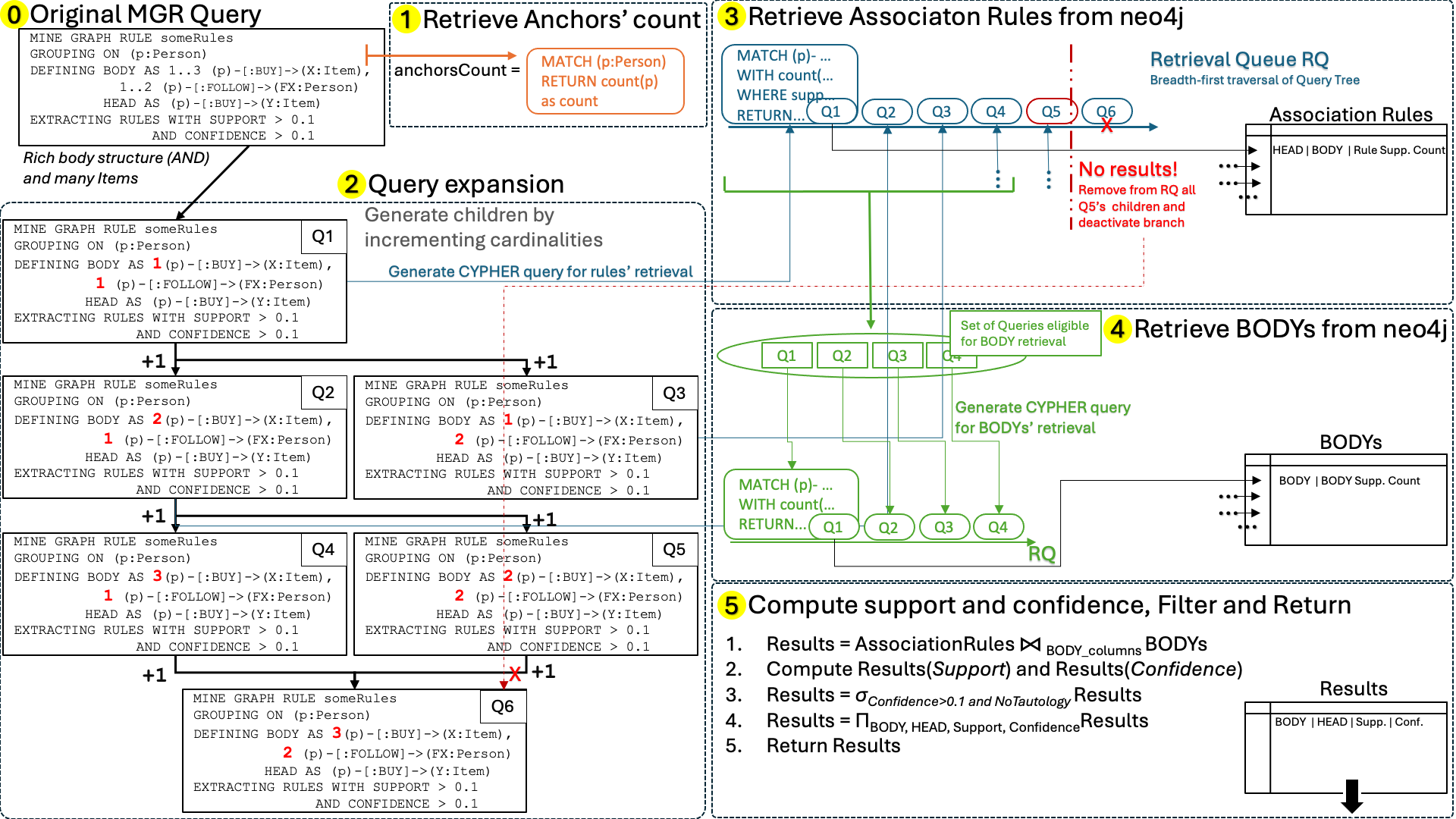}
    \caption{Phases of the \texttt{MINE GRAPH RULE} algorithm applied to an operator with two patterns in conjunction in the body, each with many items, ranging respectively from 1 to 3 and 1 to 2. This specific operator has been chosen to explain how queries are expanded, queued, and possibly removed from the queue. 
{\tt Phase(1)} counts the anchor nodes with the label `Person'. 
{\tt Phase(2)} generates various nonredundant \textit{body-head pairs}, named P1..P6, arranged in a directed acyclic graph (DAG), each with a fixed number of items in the body's conjuncts; 
for each pair, the number of items of the body's conjuncts are highlighted (in red in the figure) - while descending along the DAG levels, item numbers increase by 1, along the progression: 
[P1:(1,1)]; [P2:(2,1),P3:(1,2)]; [P4:(3,1),P5:(2,2)]; [P6(3,2)]. 
In {\tt Phase(3)}, the pairs generated at Phase (2) are queued, by performing a breadth-first traversal of the DAG; thus, the queue PQ: P1..P6 is generated. 
For each pair, a query producing head and body is issued, and the query results are entered into a table whose entries include the HEAD, the BODY, and the rule 
cardinality.
As explained in Fig.~\ref{fig:algo_base}, Phase (2), only the rules that satisfy the minimum support threshold are selected. At this stage, it is possible to remove some pairs from the queue; if we assume that no rule extracted for pair P5 satisfies the constraint on the minimum support, then P5 can be removed from the queue together with all the successors in the DAG (in this case, P6). 
In {\tt Phase(4)}, for each remaining pair P1..P4, a query producing just the body for that pair is generated, and the query results are entered into a table whose entries include the BODY and the BODY 
cardinality.
The final {\tt Phase(5)} occurs in the same way as described in Fig.~\ref{fig:algo_base}, Phase (4).}
    \label{fig:algorithm}
\end{figure*}

\subsection{Algorithm for General Cases}
\label{sub:algo_gen}

General cases include conjunctions and multiple patterns in the BODY and/or in the HEAD.
Our algorithm is inspired by the fast Apriori algorithm implementation~\cite{bodon2003fast}, where the mining starts by exploring the smaller (i.e., most common) sets of frequent items and combines them into bigger sets, until the minimum support threshold is met.
We start by considering the operators with one pattern in the BODY and one in the HEAD and then consider operators with more patterns, thus, performing a body-head pair expansion.
At each step, exactly one pattern is added either to the BODY or to the HEAD, up to the maximum number of patterns specified in the input operator. 
The body-head pairs are organized in a directed acyclic graph (DAG) from the simplest to the most complete one, as exemplified in Fig.~\ref{fig:algorithm}, Phase (2), for a particular example where the BODY presents two patterns in conjunction, the first ranging from 1 to 3 and the second ranging from 1 to 2.

Continuing our analogy with the Apriori algorithm, the objective of organizing the evaluation with a DAG of pairs is to enable the early exclusion of pairs when their support and confidence does not meet the minimum conditions set by the users. As usual, our antimonotonicity property only refers to support. 
(see Section \ref{sec:demonstration_theorem} for the related theorem and demonstration). Thus, pairs are considered in an order created by a breadth-first traversal of the DAG and, as soon as one pair does not meet the support threshold, that pair, together with all its descendants, is removed from the DAG. 
The generation of queries corresponding to each given pair is performed with the mechanism defined for simple input operators, as illustrated in Fig.~\ref{fig:algo_base}, Phases (3) and (4). 

Note that tables HEAD+BODY and BODY accumulate the instances for all the pairs. As discussed in Section~\ref{sec:algo_simple}, a final assembling step produces all the association rules with their HEAD, BODY, support, and confidence.

\subsection{Demonstration of the Antimonotonicity Property of the DAG}
\label{sec:demonstration_theorem}

To demonstrate the antimonotonicity property of the \texttt{BODY}-\texttt{HEAD} pairs of the DAG (used in the algorithm described in Section~\ref{sub:algo_gen}), we first 
introduce the DAG of BODY-HEAD pairs, where $N$ is the set of its nodes and $A$ is the set of its arcs,
then we define and prove our notion of antimonotonicity.

\begin{definition}
    Each node $n \in N$ represents a fixed \texttt{BODY-HEAD} pair of \texttt{MINE GRAPH RULE} applied to an instance $G$ of a property graph database. Each pair includes a finite set of joint patterns $P_H$ of the HEAD and a finite set of joint patterns $P_B$ of the BODY.
    For each pattern $p_{Xi} \in P_H \cup P_B$, its {\it arity} (i.e., the number of items in each conjunct) is described by the function
    $k(n, p_{Xi}): (N,P_B \cup P_H) \rightarrow \mathbb{N}^+$.
\end{definition}

\begin{definition}
\label{def:edge_dag}
    Each arc $a \in A: u \overset{a}{\rightarrow} v$ between nodes $u$ and $v$ denotes that the function $k(v, *)$ has the same output values of $k(u,*)$ for any pattern $p_{Xi}$, except for one specific pattern, denoted as $\widetilde{p_{Xi}}$, for which we impose 
    $k(v, \widetilde{p_{Xi}}) = k(u, \widetilde{p_{Xi}}) + 1$ during the construction of the DAG.
\end{definition}

\begin{theorem}
\label{th:antimonotone}
    For any node $u$ in the DAG and all its \texttt{BODY-HEAD} pair descendants $v$, the support count $S(v)$ is at most equal to the support count $S(u)$: $S(v) \leq S(u)$.    
\end{theorem}

\begin{proof}
We prove the theorem by contradiction. Suppose that for any pair $u$ such that $u\rightarrow v$, it 
holds that $S(v) > S(u)$. 
Then, the \texttt{BODY}-\texttt{HEAD} pair of $v$ is less selective than the pair in $u$, which is excluded by Definition~\ref{def:edge_dag}, imposing by construction an increased arity. We recall from the GQL semantics that any additional path -- as an added item in the result -- increases the selectivity of the \texttt{MATCH} expression. Thus, the initial assumption $S(v) > S(u)$ is false. \hfill$\square$
\end{proof}

\begin{figure*}[ht]
    \centering    
    \includegraphics[width=\linewidth]{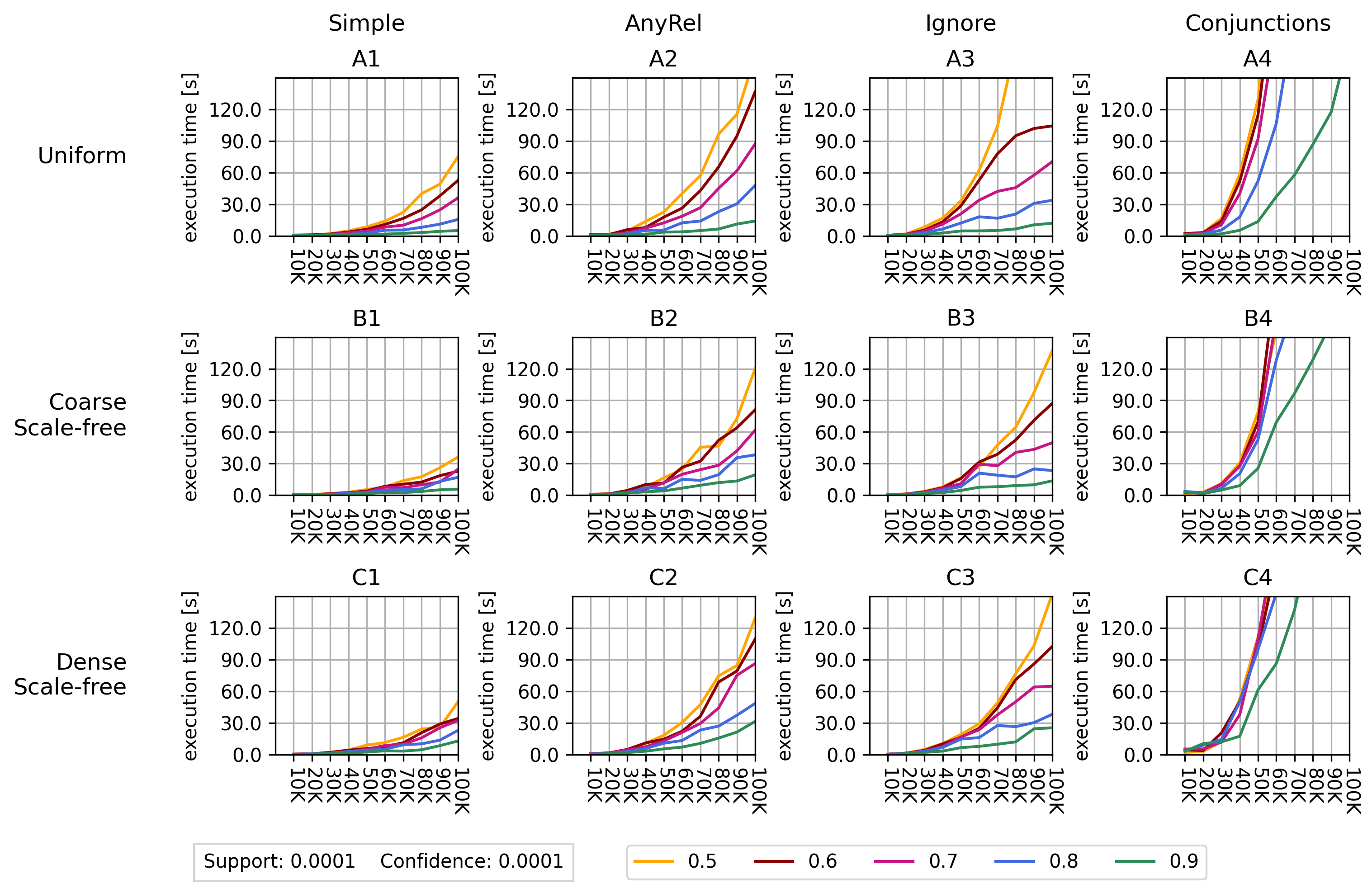}
    \caption{Performance analysis for synthetic datasets.
    Each graph reports the execution time required for executing four operators with ingresing complexity on the used patterns (respectively named Simple, AnyRel, Ignore and Many Items) over graphs with different syntethic edge generations (uniform, coarse scale-free, dense scale free); in the graph generation, we vary the total dimension of the graph (up to 100K nodes) and use five different ratios (ranging from 0.5 to 0.9) of the number of anchor nodes over the total nodes.}
\label{fig:matrix_general_scalability}
\end{figure*}

\begin{figure}[ht]
    \centering    \includegraphics[width=1\linewidth]{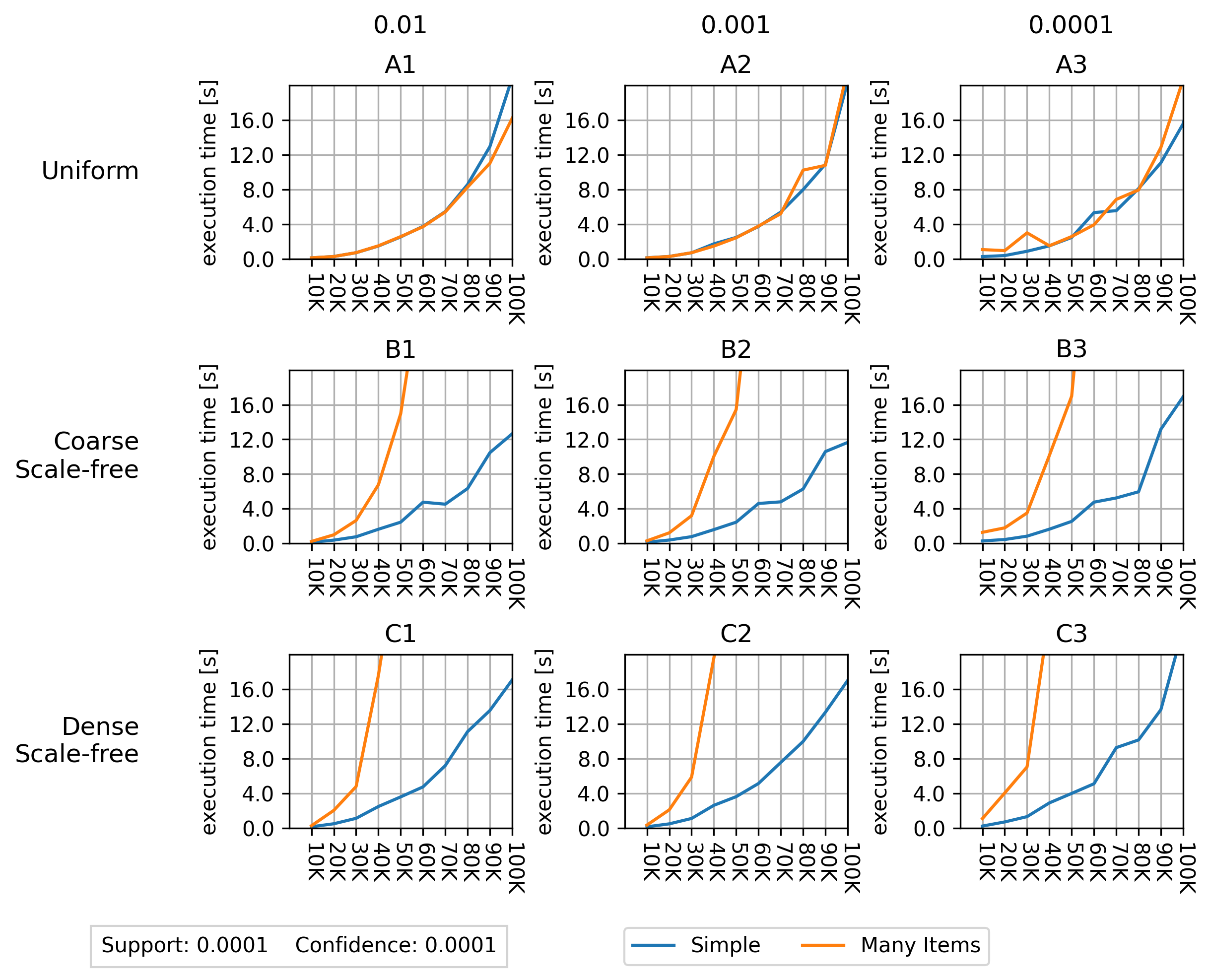}
    \caption{Impact of support in the various configurations. Each graph reports the execution time required by \textit{Simple} and \textit{Many Items} operators over graphs generated with increasing numbers of nodes, by fixing the anchor nodes over the total nodes ratio (equal to 0.8). Each column corresponds to increasing values of support (respectively 0.01, 0.001, and 0.0001); confidence is set at 0.0001. 
    }
    \label{fig:matrix_Support_scalability}
\end{figure}

\begin{figure}[!h]
    \centering
    \includegraphics[width=.8\linewidth]{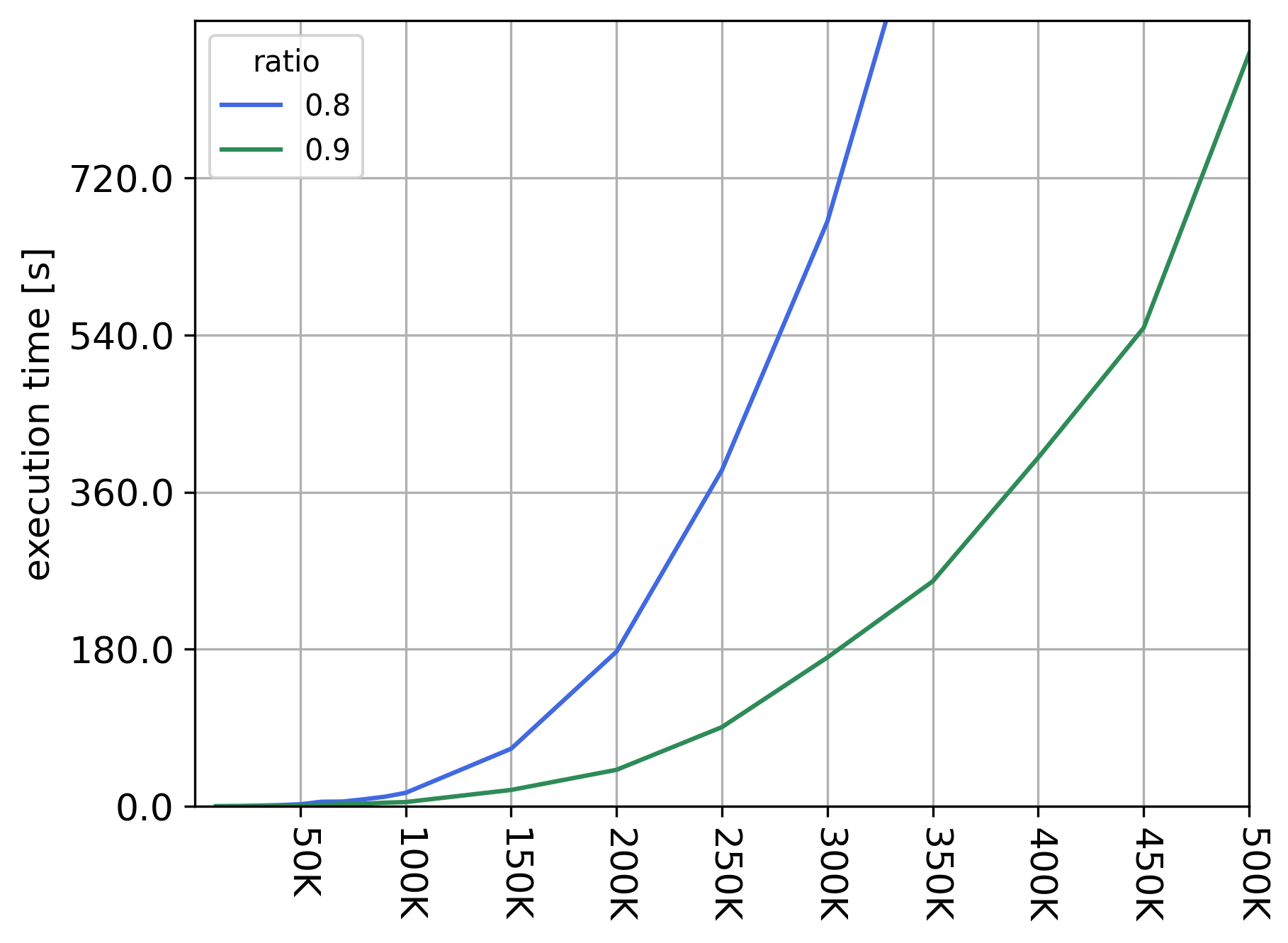}
    \caption{High Volume Scalability. For the {\it Simple} operator, we consider a high the number of nodes (up to 500K) and diferent ratios (0.8 vs 0.9) of item nodes over anchor nodes.}
    \label{fig:scalab-high_volume}
\end{figure}

\section{Evaluation}
\label{sec:evaluation}
We assessed the performance of the \texttt{MINE GRAPH RULE} operator by conducting a series of experiments over a combination of synthetic datasets and query complexities of the operator. \aggiunta{Furthermore, to validate the results obtained from the operator, we also tested it on real-world datasets.}  

\paragraph{Configuration}
\sloppy{We ran the experiments on our dedicated server machine with a 56-core Intel E5-2660 v4 CPU and 384 GB of RAM. We deployed a Neo4j database instance from its \texttt{5.15.0-community-bullseye} Docker image. To avoid memory bottlenecks, we configured the database with the following settings: }
\begin{verbatim}
server.memory.heap.initial_size=230G
server.memory.heap.max_size=230G
dbms.memory.pagecache.size=200G
\end{verbatim}

\subsection{Synthetic Databases Generation}
We generated three types of synthetic datasets, all adhering to the schema of the running example: 

\begin{itemize}
\item The first one has a {\it Uniform} distribution of all the edges among random nodes of the graph, ensuring that relationships are evenly spread. 
\item The second one, named {\it Coarse Scale-free}, uses the power-law distribution for the {\tt BUY} relationship, represents the real-case scenario of an online shop; all the other relationships are generated with a uniform distribution among all nodes. 
\item The third one, named {\it Dense scale-free}, also uses a power-law distribution for the {\tt BUY} relationship, bwith a higher density of relationships.
\end{itemize}

For each type of synthetic dataset, we generated multiple versions by varying the total number of nodes of the graph to test how the performance scales with respect to the dimensions of the graph. 
In addition, for each graph instance, we varied the ratio of anchor nodes (in our example, nodes with label {\tt Person}), which directly impacts the absolute count of {\tt BUY} relationships within a graph. 
In {\it Uniform} graphs, the total count of {\tt BUY} is proportional to the product of the number of {\tt Person} nodes and the number of {\tt Item} nodes, which is maximum when the graph is evenly split between {\tt Person}s and {\tt Item}s. Therefore, increasing the ratio over 0.5 would produce less {\tt BUY} relationships.
For the {\it Scale-free} distribution, which mimics the distribution of items' purchases in real life~\cite{mahanti2013tale,fenner2010predicting}, 
graphs have a fixed smaller portion of {\tt Item}s that are way more likely to be bought, and increasing the percentage of {\tt Person}s results in a higher number of {\tt BUY} relationships; therefore, here the ratio of anchor nodes has a nonlinear relationship with the total count of {\tt BUY} edges. 

\subsection{Experimental Results}
We evaluated the scalability of our operator in multiple settings.

\paragraph{General Scalability}
To assess the general scalability of the operator, we evaluated the execution time of four different expressions: \textit{Simple}, \textit{AnyRel}, \textit{Ignore}, and \textit{Conjunctions} referencing the examples respectively reported in Sections~\ref{sec:simpleassociationruleexample}, \ref{sec:anyassociationruleexample}, \ref{sec:ignoreassociationruleexample} and \ref{sec:and1associationruleexample}.
In Fig.~\ref{fig:matrix_general_scalability}, we compared their performance with fixed values of support and confidence (set to 0.0001) on different synthetic graphs, varying not only the distribution of relationship {\tt BUY}, but also the graph dimension (from 10K nodes to 100K nodes) and the ratios between anchor and total number nodes (from 0.5 to 0.9). In general, as the complexity of the operators grow, we obtain higher execution times while moving from Colimn 1 to Column4, as expected. 

With {\it Uniform} graphs, very few assovciation ruiles are extracted, as the topology does not encourage anomalous associations.
In {\it Scale-free} graphs, having some {\tt Item}s with higher probability to be bought results in extracting many association rules, thus activating the branch of the algorithm that would perform additional queries for conjuctive branches, ultimately making the Conjunction query slower (see Section~\ref{sub:algo_gen}).
For instance, in B3 at 50K nodes, \textit{Simple} extracts 2,630 rules while \textit{Conjuctions} extracts 10,697 rules. Thus, execution times of "Conjunctions" (cases B4/C4) are much higher tham execution times of "Simple" (cases B1/C1); performance does not seem to be too affected by the difference between dense and coarse scale-free relationship generation. 

\vspace{-3mm}
\paragraph{Impact of Support Selection}
To evaluate the effect of the support on the performance of the operator, in Fig.~\ref{fig:matrix_Support_scalability} we focused on two expressions of the operator, \textit{Simple} and \textit{Many Items} (respectively reported in Sections~\ref{sec:simpleassociationruleexample} and \ref{sec:manyassociationruleexample}), testing them with fixed values of confidence (0.0001) on synthetic graphs with different distributions and increasingly large number of nodes (from 10K to 100K). In the uniform case, as no association rules are produced, performances of Simple and Many Items are identical; instead, they differ in the scale-free case, with worse performance in the {\it Many Items} case for higher density and for reduced support costraints (as more association rules are produced in output). 

\vspace{-3mm}
\paragraph{High Volume Scalability}
Lastly, focusing on the operator \textit{Simple} (see Section~\ref{sec:simpleassociationruleexample}), we evaluated the high volume scalability with a fixed value for support and confidence (both at 0.0001) and an increasing number of nodes (up to 500K). In Fig.~\ref{fig:scalab-high_volume}, we compared two different values of ratios of anchor nodes over item nodes (0.8 in blue and 0.9 in green), showing that performances worsen when anchor nodes increase over item nodes.

\subsection{Real-World Datasets}
We evaluated the results of our operator from two real-world datasets.
\paragraph{Spotify Dataset}  \cite{spotifydataset} This dataset comprises information from over 6,000 Spotify playlists, including details such as the music genre of each playlist, the user who created it, the list of songs, the artists of the songs, and their respective genres. The resulting graph contains over 380,000 nodes and over 923,000 relationships. Consider the following {\tt MINE GRAPH RULE} operator:
\begin{Verbatim}[frame=single, commandchars=\\\{\}, fontsize=\small]
MINE GRAPH RULE ArtistPlaylist
GROUPING ON (p:Playlist)
DEFINING BODY AS (p)-[:With]->(s1:Song)
                              -[:SungBy]->(X:Artist)
         HEAD AS (p)-[:With]->(s2:Song)
                              -[:SungBy]->(Y:Artist)
WHERE X.popularity > 70 and Y.popularity > 70
IGNORE s1, s2
EXTRACTING RULES WITH SUPPORT > 0.001 
                  AND CONFIDENCE > 0.001 
\end{Verbatim}
The interpretation of the generated association rules is {\it ``Playlists with songs sung by X (with over 70 popularity), have also songs sung by Y (with over 70 popularity)''} and with these thresholds values for support and confidence the operator extracts more than 60,000 rules. Table~\ref{tab:ArtistPlaylistTable} reports the rules with the 10 highest support values. The top positions are mainly occupied by paired rules (i.e., where the artists are swapped between the head and body). 
Based on the mathematical definition of support, these paired rules must have the same support value. 
However, their confidence values can vary significantly, which suggests interesting implications. 
For instance, {\it ``Playlists with songs sung by Harry Styles, have also songs sung by Taylor Swift''} has support of 0.0181 and confidence of 0.491, while {\it ``Playlists with songs sung by Taylor Swift, have also songs sung by Harry Styles''} has the same support of 0.0181 but confidence of 0.348, suggesting that Harry Styles fans enjoy Taylor Swift songs more than her fans enjoy his songs. This can be interpreted as a greater attention to Taylor's Swift's personal history, which is typically in the text of her songs. 
\begin{table}[h!]
  \centering
  \resizebox{\columnwidth}{!}{
  \begin{tabular}{cccc}
    \toprule
    \textbf{Body} & \textbf{Head} & \multirow{2}{*}{\textbf{Support}} & \multirow{2}{*}{\textbf{Confidence}}\\ 
    \cmidrule(r){1-1}\cmidrule(l){2-2}
    \textbf{PlaylistWithSongSungByX} & \textbf{PlaylistWithSongSungByY} & &\\
    \midrule 
    Kanye West & Drake & 0.0216 & 0.475 \\ 
    Drake & Kanye West & 0.0216 & 0.446 \\
    Kendrick Lamar & Drake & 0.0204 & 0.582 \\ 
    Drake & Kendrick Lamar & 0.0204 & 0.422 \\ 
    21 Savage & Drake & 0.0183 & 0.816 \\
    Drake & 21 Savage & 0.0183 & 0.377 \\ 
    Harry Styles & Taylor Swift & 0.0181 & 0.491 \\ 
    Taylor Swift & Harry Styles & 0.0181 & 0.348 \\ 
    Kendrick Lamar & Kanye West & 0.0176 & 0.502 \\ 
    Drake & Future & 0.0176 & 0.364 \\ 
    \bottomrule
  \end{tabular}
  }
  \caption{Output of ArtistPlaylist rules with highest 10 values of support}
  \label{tab:ArtistPlaylistTable}
\end{table}

\paragraph{ArXiv Dataset}  \cite{arxiv_org_submitters_2024} This dataset comprises information from scientific papers, including details on both their authors and their categories/macro-categories. The resulting graph contains over 5 million nodes and over 15 million relationships. Consider the following \texttt{MINE GRAPH RULE} operator:
\begin{Verbatim}[frame=single, commandchars=\\\{\}, fontsize=\small]
MINE GRAPH RULE AuthorCategory
GROUPING ON (a:Author)
DEFINING BODY AS (a)-[:Publish]->(ar1:Article)
                        -[:Labelled]->(CX:Category)
                        -[:Of]->(m1:Macrocategory)
                        AS "ComputerScience"
         HEAD AS (a)-[:Publish]->(ar2:Article)
                        -[:Labelled]->(CY:Category)
                        -[:Of]->(m2:Macrocategory)
                        AS "Physics"
WHERE m1.name = "Computer Science"
        and m2.name = "Physics"
IGNORE ar1, ar2, m1, m2
EXTRACTING RULES WITH SUPPORT > 0.001 
                  AND CONFIDENCE > 0.001 
\end{Verbatim}
The generated association rules can be interpreted as {\it ``Authors who publish articles labelled CX of Computer Science, also publish articles labelled CY of Physics''}. In this case, we opted for simpler column names (using the aliases), and named columns just {\it ComputerScience} and {\it Physics}).
The operator extracted over 1,000 rules using these threshold values for support and confidence, demonstrating its capability to handle millions of nodes, and the rules with the highest 10 values of support are listed in Table~\ref{tab:AuthorCategoryTable}. For instance, the pair {\it Machine Learning, Quantum Physics} has support of 0.0042 and confidence of 0.039. 

\begin{table*}[h!]
  \centering
  \resizebox{1.4\columnwidth}{!}{
  \begin{tabular}{cccc}
    \toprule
    \textbf{Body} & \textbf{Head} & \multirow{2}{*}{\textbf{Support}} & \multirow{2}{*}{\textbf{Confidence}}\\ 
    \cmidrule(r){1-1}\cmidrule(l){2-2}
    \textbf{ComputerScience} & \textbf{Physics} & &\\
    \midrule 
    Machine Learning & Material Science & 0.0042 & 0.040 \\ 
    Machine Learning & Quantum Physics & 0.0042 & 0.039 \\
    Machine Learning & Computational Physics & 0.0032 & 0.030 \\ 
    Computer Vision and Pattern Recognition & Materials Science & 0.0031 & 0.040 \\ 
    Artificial Intelligence & Materials Science & 0.0029 & 0.040 \\
    Machine Learning & Optics & 0.0029 & 0.027 \\ 
    Machine Learning & Mesoscale and Nanoscale Physics & 0.0029 & 0.027 \\ 
    Artificial Intelligence & Quantum Physics & 0.0028 & 0.038 \\ 
    Social and Information Networks & Physics and Society & 0.0027 & 0.196 \\ 
    Machine Learning & Physics and Society & 0.0027 & 0.025 \\ 
    \bottomrule
  \end{tabular}
  }
  \caption{Output of AuthorCategory rules with highest 10 values of support}
  \label{tab:AuthorCategoryTable}
\end{table*}


\section{Brief Comparison with Path Association Rules Mining (PARM)}
\label{sec:comparison}

Among the graph pattern mining algorithms in the existing literature, PARM, by Sasaki and Karras~\cite{sasaki2024mining} (Nov. 2024) exploits some of the characteristics of property graphs to extract association rules. 
This section provides a deep analysis of the similarities and differences between PARM and our approach, highlighting their respective strengths.

PARM explores association rule mining by leveraging paths within the graph; path patterns are classified either as 
\textit{simple paths}, i.e. paths of a given length consisting of alternating node labels (attributes) and edge types, or 
\textit{reachability path patterns}, where only the labels of the first and last nodes are specified, and these nodes are connected by a path of edges of the same type, constrained to a maximum length. 
PARM defines a path as matching a path pattern when it is composed exactly of the alternating sequence of nodes and edges specified in the pattern; it also states that a node matches a path pattern if it originates a path that matches to the same pattern. PARM also includes the dominance between pairs of path patterns, stating that one pattern dominates another if it is longer and matches the other pattern along its entire length.
These premises allow the definition of Path Association Rules (PAR) as pairs of path patterns, as well as various notions of support and confidence. In particular, the authors focus on association rules between paths with support above a given threshold, shorter than a maximum length, and such that neither of them dominates the other. Rule extraction is performed by the PIONEER algorithm, which takes advantage of antimonotonicity to prune infrequent patterns and of auxiliary data structures to store intermediate results; several extensions of the base version of the method are discussed, for approximation and parallelization.  

In comparison, while \texttt{MINE GRAPH RULE} provides syntax and semantics for declaratively describing association rules, PARM extracts association rules from the graph in a bottom-up manner. 
A few aspects are similar, e.g., the use of path expressions; 
\texttt{MINE GRAPH RULE} does not support reachability paths but it supports the orthogonal and recursive construction of patterns with much higher expressive power, as these include aggregation, alternative paths, arbitrary composition of itemSets, conjunctive expressions of paths, and arbitrary predicates over any semantic feature supported by property graphs (e.g., arbitrary labels and properties of nodes and edges) -- thanks to the richness of GQL. Algorithms are not comparable, as PIONEER is implemented in C++ over structures optimized for the purpose.

\section{Discussion and Conclusion}
\label{sec:conclusion}

After our progressive illustration of the expressive power of the operator in Section~\ref{sec:examples}, we summarize here the benefits of our approach in terms of readability.  
\begin{itemize}
\item Our language uses clauses mimicking GQL expressions, employing standard GQL variables and predicates over properties and labels of entities and relationships; as such, it exploits the richness of the property graph data model. 
\item Resulting association rules are extracted as rows of (structured) tables, associating the Body and Head itemSets to their support and confidence. These can be inspected by any non-technical user in the form of a CSV file, e.g., by using spreadsheets.
\item We also developed a recursive mechanism, driven by a grammar, to generate semantically rich column names, e.g., for Body: \textbf{FollowPersonRecommendItemOfCX} and for Head \textbf{BuyItemOfCY} (see Table~\ref{tab:MoreComplexBodyAssociationRulesTable}); when names are considered as too complex, they can be renamed using alias clauses (see Table \ref{tab:AuthorCategoryTable}).
\end{itemize}
Some recent papers have attempted to translate complex formal expressions into readable examples, such as {\it two users are likely colleagues if they follow the same organization and have over $k$ friends in common} (see Example~1 in  \cite{fan2022parallel}) or {\it if a television station v is part of a company employing a CEO, then it is also part of a company employing a professor}  (see Example `Nell' in~\cite{sasaki2024mining}). However, these rules are not equipped with their support and confidence, and no indication is given about how to extract these specific rules from the large collection of extracted rules. 

Our approach exploits the use of a GQL engine; in particular, the fragment of GQL that we use is also compatible with Cypher~\cite{francis2018cypher}, a widely used graph database language. Therefore, in our implementation, we map \texttt{MINE GRAPH RULE} patterns to Cypher and execute them on a Neo4j engine; we use the Apriori approach at a set-oriented level, and exploit the query optimizer of the graph database engine, without resorting to schema-specific tunings or heuristics. 
Although our prototype implementation is specific to Neo4j, the syntax and semantics of the operator are general, and the principles followed in the implementation can be applied to any graph database. 

In this work, our focus has been on defining the syntax and semantics of the {\tt MINE GRAPH RULE} operator; we deliberately left out the optimization of the execution over a data mining engine. Clearly, the \textit{a priori} knowledge of the queries makes them amenable to efficient optimizations. For what concerns query expansion, we could design a special index that expands one group if and only if it has enough support, mimicking with a physical structure an optimized breadth-first DAG traversal.
We intend to dedicate follow-up work to deepening query optimization for property graph databases, using suitable physical database ingredients, and performance-oriented benchmarks.

In the current design, {\tt MINE GRAPH RULE} is a declarative, top-down operator; another interesting direction for future work is to embed the operator into higher-level abstractions, so as to give more space to bottom-up identification of most frequent patterns.

In conclusion, \texttt{MINE GRAPH RULE} is a significant step forward along the directions discussed in~\cite{bonifati2025roadmap}, to which we recently contributed in~\cite{invernici2024searching,colombo2025legislative,colombo2025llm}.



\section*{Resources}
The source code of the implementation of the mineGraphRule plugin for Neo4j, examples of usage, and the scripts to generate the evaluation datasets are available on GitHub \cite{cambria2024repo-ORIGINAL}. 

\begin{acknowledgements}
This paper is supported by the FAIR (Future Artificial Intelligence Research) project, funded by the NextGenerationEU program within the PNRR-PE-AI scheme (M4C2, Investment 1.3, Line on Artificial Intelligence).
\end{acknowledgements}

\bibliographystyle{acm}
\bibliography{vldb-bibliography}   

\balance

\end{document}